\documentclass[sigconf, 10pt]{acmart}
\usepackage{epsfig}
\usepackage{url}
\usepackage{color}
\usepackage{graphicx}
\usepackage{caption}
\usepackage{amsfonts}
\usepackage{algorithm}
\usepackage{amsmath}
\usepackage{float}
\usepackage{multirow}
\usepackage{extarrows}
\usepackage{comment}
\usepackage{booktabs}
\usepackage{tabularx}
\usepackage{diagbox}
\usepackage{balance}
\usepackage{calc}
\usepackage{textcomp}
\usepackage{subcaption}
\usepackage{tikz}
\usepackage{hyperref}

\DeclareMathDelimiter{(}{\mathopen} {operators}{"28}{largesymbols}{"00}
\DeclareMathDelimiter{)}{\mathclose}{operators}{"29}{largesymbols}{"01}

\makeatletter

\newcommand \SysName {G-Tran}

\newcommand{\tabincell}[2]{\begin{tabular}{@{}#1@{}}#2\end{tabular}}
\newcolumntype{P}[1]{>{\centering\arraybackslash}p{#1}}

\def\hlinew#1{
	\noalign{\ifnum0=`}\fi\hrule \@height #1 \futurelet
	\reserved@a\@xhline}

\makeatother

\AtBeginDocument{%
  \providecommand\BibTeX{{%
    \normalfont B\kern-0.5em{\scshape i\kern-0.25em b}\kern-0.8em\TeX}
	}
}

\setcopyright{acmcopyright}
\renewcommand\footnotetextcopyrightpermission[1]{}
\begin{document}

\title{G-Tran: Making Distributed Graph Transactions Fast}

\author{Hongzhi Chen, Changji Li, Chenguang Zheng, Chenghuan Huang, Juncheng Fang, James Cheng, Jian Zhang}
\affiliation{Department of Computer Science and Engineering \\ The Chinese University of Hong Kong}
\email{{hzchen, cjli, cgzheng, chhuang, jcfang6, jcheng, jzhang}@cse.cuhk.edu.hk}

\renewcommand{\shortauthors}{Hongzhi Chen et al.}
\renewcommand{\textrightarrow}{$\rightarrow$}

\begin{abstract}
Graph transaction processing raises many unique challenges such as random data access due to the irregularity of graph structures, low throughput and high abort rate due to the relatively large read/write sets in graph transactions. To address these challenges, we present \SysName{} --- an RDMA-enabled distributed in-memory graph database with serializable and snapshot isolation support. First, we propose a graph-native data store to achieve good data locality and fast data access for transactional updates and queries. Second, \SysName{} adopts a fully decentralized architecture that leverages RDMA to process distributed transactions with the MPP model, which can achieve high performance by utilizing all computing resources. In addition, we propose a new MV-OCC implementation with two optimizations to address the issue of large read/write sets in graph transactions. Extensive experiments show that \SysName{} achieves competitive performance compared with other popular graph databases on benchmark workloads.

\end{abstract}

\maketitle

\section{Introduction}\label{intro}
Graph data are abundant today in both industrial applications and academic research. In order to support efficient graph data storage and management, graph databases have become an essential infrastructure. However, most existing graph databases~\cite{titan, janusgraph, orientdb, arangodb, neo4j, tao, TigerGraph, Weaver, geabase, KyrolaG14} have shortcomings in their design and functionality that lead to performance bottlenecks in processing complex graph transactions. For example, JanusGraph~\cite{janusgraph}, OrientDB~\cite{orientdb} and ArangoDB~\cite{arangodb} use BigTable~\cite{ChangDGHWBCFG06} or Multi-Model storage to represent graphs, leading to significant read/write amplifications during graph queries and updates. Neo4j~\cite{neo4j} and TigerGraph~\cite{TigerGraph} call themselves \textit{native graph stores}, as to distinguish from the non-native stores (e.g., relational, columnar, key-value stores), and represent graphs as adjacency-lists or compressed formats (e.g., CSR, CSC). These data layouts are more efficient for graph queries but suffer from poor data locality when updates are frequent, thus incurring extra overheads on data retrieval and impairing both latency and throughput (\S\ref{moti}).

In addition to handling graph-specific read/write, transaction isolation is another important functionality in DBMSs, which controls how transaction integrity is visible to concurrent users in order to maintain the correctness of transaction processing. Transaction isolation (ideally strict serializability) is challenging when dealing with large-scale graph data for the following reasons. Read-only graph transactions tend to have large \textbf{read sets} since a large number of vertices and edges can be easily involved after just two to three hops of traversal starting from a vertex, as most real-world graphs exhibit a power-law degree distribution. Similarly, read-write transactions may also have large \textbf{write sets} (see Figure~\ref{mixed-size}). These large read/write sets lead to high contention in concurrent transaction processing. As graph transactions have relatively long processing time, consequently the contention becomes more serious, which in turn leads to high abort rate and low throughput.

These unique challenges in graph transaction processing motivate us to design a new distributed graph database system, called \textbf{\SysName{}}, for high-performance transaction processing on property graphs~\cite{AnglesG08}. To the best of our knowledge, \SysName{} is the first RDMA-enabled graph database that provides strong consistency, i.e., \textit{serializability} (\textit{SR}) and \textit{snapshot isolation} (\textit{SI}), low latency and high throughput for graph transaction processing. We highlight some unique designs of \SysName{} as follows:

\vspace{-1mm}
\begin{itemize}
	\setlength{\parskip}{0pt}
	\setlength{\itemsep}{0pt plus 1pt}
	
	\item We design a \textit{graph-native} data store with efficient data and memory layouts, which offers good data locality and fast data access for read/write graph transactions under frequent updates.
	
	\item We propose a \textit{fully decentralized system architecture} by leveraging the benefits of RDMA to avoid the bottleneck from centralized transaction coordinating, and each worker executes distributed transactions under the MPP (i.e., massive parallel processing) model.
	
	\item \SysName{} presents a \textit{multi-version-based optimistic concurrency control} (\textit{MV-OCC}) protocol, which is specifically designed to reduce the abort rate and CPU overheads in concurrent graph transaction processing.
	
	\vspace{-1mm}
\end{itemize}

We demonstrate the effectiveness of our system designs and the overall performance by comparing \SysName{} with the state-of-the-art graph databases~\cite{janusgraph, arangodb, neo4j, TigerGraph} using benchmark workloads~\cite{LissandriniBV18,ErlingALCGPPB15}. The results show that \SysName{} can achieve up to orders of magnitude improvements over the existing graph databases and obtain high throughput at both SR and SI isolation levels.
\section{Background}\label{sec:bg}
\noindent \textbf{Property Graph.}	\ \SysName{} adopts the \textit{property graph} (\textit{PG}) model to represent graph data because of the generality and expressiveness of PG. In a PG, vertices represent entities in an application and (directed) edges model the relationships between two entities. Both entities and relationships may have a set of properties to describe their attributes (e.g., names, gender, etc.) in the form of key-value pairs. Many existing graph databases, such as Titan~\cite{titan}, JanusGraph~\cite{janusgraph}, OrientDB~\cite{orientdb} and Neo4J~\cite{neo4j}, adopt PG to model their graph data, while using \textit{Gremlin}~\cite{gremlin} as the query language. Currently, \SysName{} also uses Gremlin (the latest 3.0 standard) as its query language.

\vspace{1mm}
\noindent \textbf{Concurrency Control Protocols.} \ Concurrency control ensures atomicity and isolation for database transactions. \SysName{} supports both serializability (SR) and snapshot isolation (SI). SR has the strictest constraint that all concurrent transactions should execute their operations logically as if they are executed in sequence. SI relaxes the constraint to require that only all reads in a transaction should see a consistent \textit{snapshot} of the database. There are three kinds of concurrency control protocols that are widely used in databases to implement different isolation levels: \textit{two-phase locking} (\textit{2PL})~\cite{EswarranGLT76}, \textit{optimistic concurrency control} (\textit{OCC})~\cite{KungR79}, and \textit{multiple version concurrency control} (\textit{MVCC})~\cite{Reed78}. 2PL is the most common and simplest protocol, which uses locks to avoid conflicts among concurrent transactions. OCC does not use lock, but it avoids conflicts by validation after a transaction completes its execution. Generally, OCC handles transactions in three steps: \textit{process}, \textit{validate}, and \textit{commit/abort}. In comparison, MVCC provides \textit{point-in-time consistent views} for multiple transactions at the same time by maintaining multi-versions of each object with timestamps, which incurs a higher storage overhead.


\vspace{1mm}
\noindent \textbf{InfiniBand and RDMA.} \ InfiniBand has become quite commonly in use recently and led to the development of many new distributed systems~\cite{Sahuquillo18, IslamWJRWSMP12, KamburugamuveRS17, ZhangLP17}. InfiniBand offers two network communication stacks: IP over InfiniBand (IPoIB) and \textit{remote direct memory access} (\textit{RDMA}). IPoIB implements a TCP/IP stack over InfiniBand to allow current socket-based applications to be executable without modification. In contrast, RDMA provides a \textit{verbs} API, which enables zero-copy data transmission through the \textit{RNIC} without involving the OS. RDMA has two verb operations: two-sided \textit{send}/\textit{recv} verbs and one-sided \textit{read}/\textit{write} verbs. Two-sided verbs provide a socket-like message exchange mechanism, which still incurs CPU overheads on remote machines. One-sided verbs can directly bypass both the kernel and CPU of a remote machine to achieve low latency.
\section{Challenges and Design Choices}\label{moti} 

We have briefly discussed the characteristics of graph transaction processing and the limitations of existing graph databases in~\S\ref{intro}. In this section, we analyze the important factors that should be considered in the design of a high-performance distributed graph database. We summarize the challenges of distributed graph transaction processing as follows.

(C1)~Graph data can easily result in poor locality for reads and writes in databases after continuous updates due to its irregularity~\cite{sun2015sqlgraph}.

(C2)~Due to the power-law distribution on vertex degree and the small world phenomenon of most real-world graphs, the cost of traversal-based queries can be very high after multi-hops (e.g., $\geq$ 2) fan-out~\cite{ShalitaKKSPAKS16}.

(C3)~Graph traversal query can involve relatively larger read/write set and longer execution time compared with those queries on other storage models (e.g., KVS) due to the connectedness of graph data, which may lead to higher contention and lower throughput in transactions.

(C4)~Latency and scalability is another critical issue for distributed graph transactions. The network bandwidth, contention likelihood, and CPU overheads are the main bottlenecks~\cite{Myth}. Traditional centralized system architecture (i.e., assigning a master node as the global coordinator) may limit the scalability for OLTP.

\begin{figure*}[!t]
	\centering
	\includegraphics[width=0.82\textwidth]{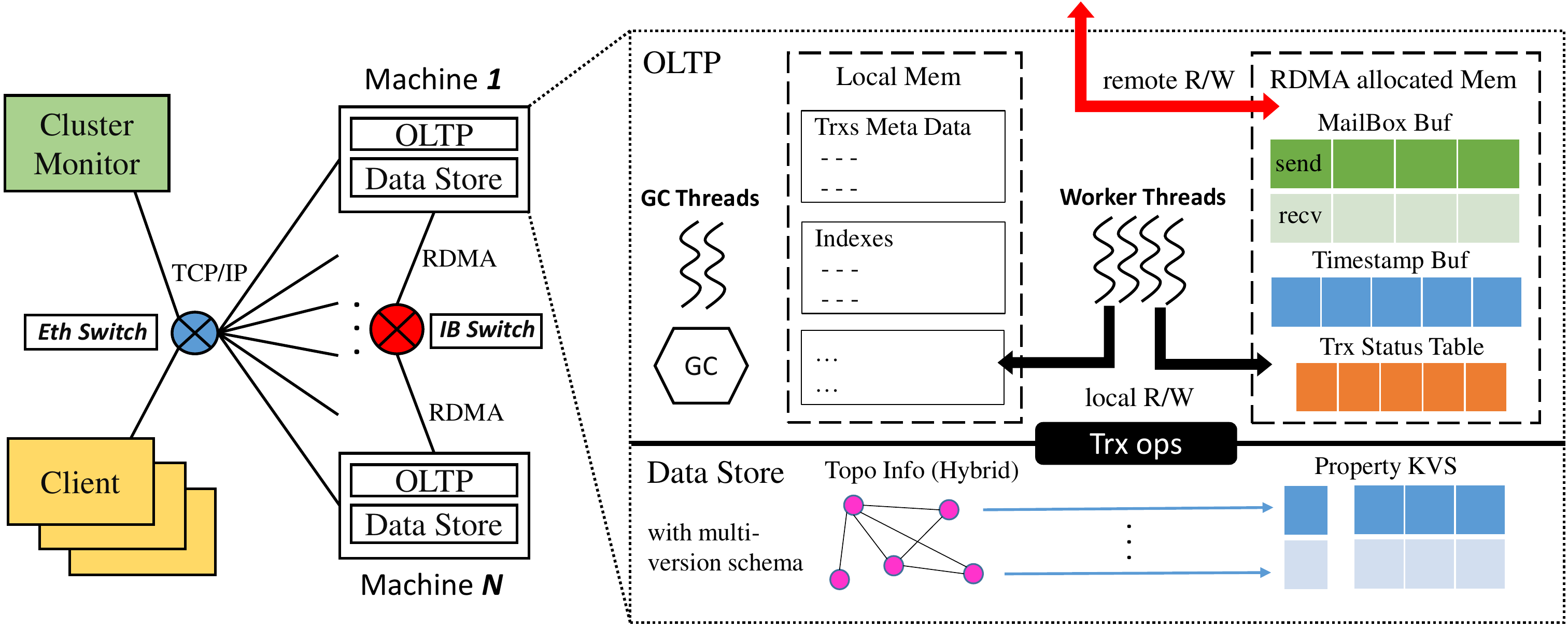}
	\vspace{-3mm}
	\caption{The architecture overview of \SysName{} (best viewed in color)}\label{arct}
	\vspace{-2mm}
\end{figure*}

\vspace{1mm}
We conduct experiments in~\S\ref{exp} to verify the above issues. Here, we discuss the design choices to address the aforementioned challenges.

We first consider (C4). A significant overhead for distributed transaction processing is the large number of round-trip messages that are needed to ensure ACID. In recent years, many systems~\cite{binnig2016end, ChenWSCC16, NNRSB15, KaliaKA14, MitchellGL13, RNCDNC19, WeiSCCC15, Myth} have been proposed to improve the performance of distributed transaction processing by leveraging RDMA as it can remove the overheads on network and CPU, although these works mainly focus on RDBMS or KVS instead of graph databases. Using RDMA for graph transaction processing can also improve the performance in terms of throughput, latency and scalability. However, the design of an RDMA-based distributed in-memory database is non-trivial. Specifically we move from a pure shared-nothing or shared memory architecture to a hybrid memory layout (described as below) via RDMA one-sided read/write and two-sided send/recv. These new features require us to have an overall system redesign from the storage layer to the transaction processing layer in order to tightly integrate each system component.

To address (C1) and (C2), we propose a new storage layout for property graphs to achieve both good data locality and efficient data access. It stores graph data under a graph-native schema, but organizes the (arbitrary-length) adjacency-lists of vertices into fixed-size rows and separately maintains vertex/edge properties into a key-value store. The storage layer of \SysName{} splits the memory space of each machine into two parts. One part follows the shared-nothing architecture to store one piece of graph partition locally, and the other part follows the shared-memory architecture to compose an RDMA-based distributed memory space for remote property data access. Such design is based on the facts that: (1)~to implement efficient RDMA-based data structures (e.g., map, tree) are non-trivial, as the address space of each object should be recorded explicitly and this requires more memory footprint~\cite{0001VBFK19}; (2)~not all data are necessary to be shared and to manage different regions of data in different memory spaces compactly can further improve the data access efficiency for local search and scan (\S\ref{store}).

For (C3), 2PL would incur high overhead due to locking on large portions of data. OCC is a good choice for read-intensive workloads, but we can further reduce the overhead of isolation maintenance in concurrent transactions by integrating OCC with multi-versioning. We adopt a \textit{multi-version OCC} (\textit{MVOCC}) protocol to coordinate concurrent transaction processing. MV-OCC has the following advantages. First, a multi-versioning mechanism allows read-only transactions to access old versions of objects without imposing any consistency overhead on read-write transactions. Second, OCC applies validation to check if there is any conflicting updates without locks. We propose our own optimistic optimizations in the MVOCC protocol to reduce the abort rate (\S\ref{proto}).

Moreover, \SysName{} adopts an MPP-enabled decentralized architecture to process transaction execution (\S\ref{exec}). Existing distributed databases~\cite{NNRSB15, Weaver} follow a centralized approach to process tasks such as transactional metadata management and timestamp ordering on a master node. The master is likely a bottleneck of the entire cluster and affects the overall performance~\cite{Myth}. Instead, in our decentralized architecture, all servers share the load of each transaction according to the locality of its read/write sets and each server can handle multiple sub-queries simultaneously with multi-threads (i.e., MPP) for speed-up. In addition, RDMA can significantly reduce the costs of essential operations such as clock synchronization, timestamp ordering, transaction status synchronization in the decentralized architecture, which also helps improve the scalability.
\section{System Design}\label{design}

\begin{figure*}[!t]
	\centering
	\includegraphics[width=0.9\textwidth]{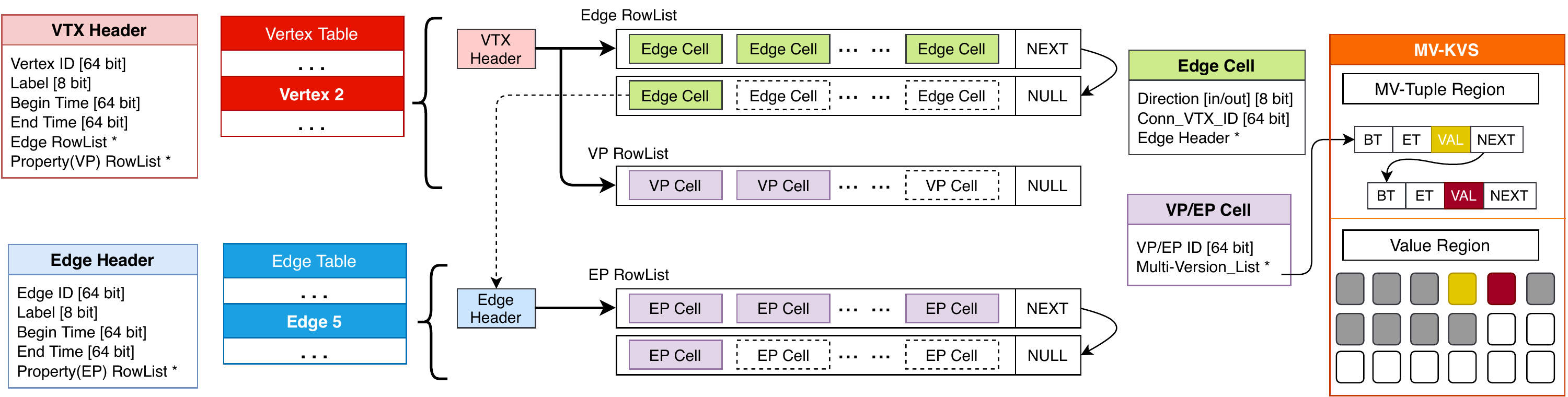}
	\vspace{-3mm}
	\caption{The data store of \SysName{} (best viewed in color)}\label{datastore}
	\vspace{-2mm}
\end{figure*}

\subsection{Overview}\label{arch}
Figure~\ref{arct} shows the architecture of \SysName. \SysName{} supports multiple client connections to server nodes through a regular Ethernet network. Each client provides a user console and a light communication lib for submitting transactions and receiving query results. A cluster monitor monitors the load of each server by a heartbeat mechanism and conducts the assignment of incoming transactions from clients for load balancing among servers. \SysName{} server nodes are connected with each other via InfiniBand for RDMA communication. Each server node is composed of two layers, i.e., the storage layer  and the OLTP layer.

\vspace{1mm}
\noindent  \textbf{The storage layer}, i.e., the data store, keeps a property graph in two parts: \textit{topology data} and \textit{property data}. Topology data refer to the graph topology, i.e., vertices and edges in the format of adjacency lists. Property data are the keys and values of the vertex/edge properties. We partition a graph into $N$ shards over $N$ server nodes. \SysName{} constructs a consistent global address space over all the nodes in a cluster, so that the location of any object in the data store can be retrieved by its ID. The data store has a multi-version schema to support consistent view for concurrent transactions (\S\ref{store}).

\vspace{1mm}
\noindent  \textbf{The OLTP layer} of each server node sets a group of worker threads to process incoming transactions. Worker threads interact with the data store to process data reads/writes. To allow remote data access and fast communication via RDMA, each server node registers a chunk of memory at NIC during the initialization stage, which divides the memory space of a server node into two regions, \textit{RDMA allocated memory} and \textit{local memory}, as depicted in Figure~\ref{arct}. We place different system components in different memory regions according to their functionalities, in order to enjoy the benefits of both local memory management (i.e., efficient maintenance and access of data structures such as tree, map, lock, etc.) and RDMA (i.e., low CPU overhead, fast remote data access and atomicity guarantee) for concurrent transaction processing.

\SysName{} adopts MPP model~\cite{TigerGraph} in the OLTP layer (\S\ref{exec}). That is, a cross-server transaction is executed simultaneously in the places where its data (i.e., the read/write sets) are located and each transactional operation can be processed in parallel based on its load. The MPP model leads to better data locality and enables parallel execution of a single transaction. As a trade-off, parallel execution over multiple servers requires more message passing for consistency control, but we keep the network communication overhead low by using RDMA. We construct an RDMA-enabled mailbox (i.e., the green box in Figure~\ref{arct}) to process point-to-point message sending/receiving at thread-level, including one-sided RDMA read/write and two-sided RDMA send/recv. Besides the mailbox, in the RDMA allocated memory, \SysName{} also maintains some system components (e.g., \textit{timestamp bufs}, \textit{transaction status table}) for global transactional metadata access, which is needed in our MV-OCC protocol (\S\ref{proto}). However, the transactions' private metadata (e.g., begin time, commit time, read/write set) and other structures (e.g., indexes) that do not need to be shared with other servers, are stored in the local memory.

\subsection{Data Store}\label{store}
As shown in Figure~\ref{datastore}, the data store in each server is composed of three components: \textit{Vertex Table} (in red color), \textit{Edge Table} (in blue color), and \textit{Multi-Version Key-Value Store} (\textit{MV-KVS}, in orange).

All vertex objects are stored in the Vertex Table. Every vertex object has a fixed-size \textit{vertex header} to record the information (e.g., ID and label) of a vertex. The \textit{begin time} and \textit{end time} in the vertex header indicate the \textit{visible time period} of the vertex, i.e., the period that the vertex is accessible to all active transactions. Usually, the begin/end time of a vertex is exactly the commit time of the transaction that creates/deletes this vertex. In addition, the vertex header also links to two \textit{row-lists}, which record the connected edges (i.e., \textit{Edge RowList *}) and properties (i.e., \textit{VP RowList *}) of the vertex, respectively. To store the adjacency list of a vertex, we use an \textit{Edge Cell} (in green) to represent each adjacent vertex, and then arrange all the Edge Cells in an adjacency list into rows in ascending order of the vertex IDs. Note that each row has a fixed number of cells. If one row is filled up, a new row will be allocated from the memory pool until the entire adjacency list is stored.

Each edge cell is mapped to one edge object stored in the Edge Table. Since an edge $e=$($v_1,v_2$) connects two vertices,  there are two edge cells (of $v_1$ and $v_2$) pointing to the same edge object of $e$. If $e$ is directed, then the edge cell of $v_1$/$v_2$ also keeps a direction sign to indicate that $e$ is an out/in-edge of $v_1$/$v_2$. Each edge object also has an \textit{edge header}, which records its ID, label, begin time, end time, and a link to the row-list, i.e., \textit{EP RowList *}, that stores the properties of the edge. Both VP RowList and EP RowList have the same layout as that of Edge RowList. Each \textit{VP/EP Cell} in the VP/EP RowList records the ID of a property object and a pointer that links to its multi-version property values in the MV-KVS.

\begin{figure*}[!t]
	\centering
	\includegraphics[width=0.88\textwidth]{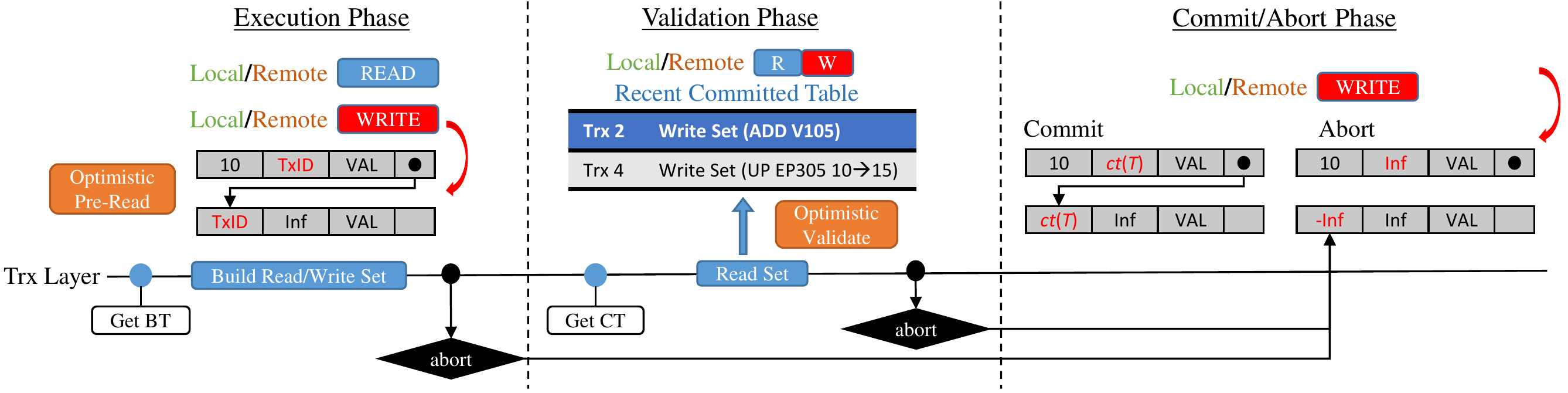}
	\vspace{-3mm}
	\caption{The MV-OCC protocol of \SysName}\label{mvocc}
	\vspace{-2mm}
\end{figure*}

The MV-KVS is divided into two regions as shown in Figure~\ref{datastore}:  \textit{MV-tuple region} and \textit{value region}. The MV-tuple region stores a set of pre-allocated, fixed-size MV-tuples, where each MV-tuple records the begin and end time of a version of the corresponding property. All MV-tuples (i.e., the different versions) of a property object are ordered by their begin/end time for version searching during data access. Each MV-tuple keeps a pointer that links to where the value of this version is stored in the value region. We will discuss how to operate on the multi-versions of a property object in~\S\ref{proto}.


\vspace{1mm}
\noindent  \textbf{Design Principle.} The key design of our data store is that all components have fixed sizes (except property values with variable lengths) and are aligned compactly in memory space wherever possible. This provides good data locality and is critical for efficient read and write (e.g., insert a new edge). As dynamic memory allocation is a costly operation, \SysName{} uses memory pooling to pre-allocate memory buffers for all type of components (i.e., headers, cells, row-lists, tuples). This data layout enables that graph traversals will be executed as a combination of row-based scanning and object-based filtering (on MV-KVS) via \textit{zero-CPU-overhead} one-sided RDMA read/write. In addition, it also benefits the multi-threaded execution of concurrent transactions (\S\ref{exec}).

\subsection{The MV-OCC Protocol}\label{proto}
To guarantee a transaction $T$ is serializable, we should hold two features in a multi-version storage. (1)~\textit{Read stability}: the visible version of a record for $T$ should keep unchanged during the processing of $T$. Traditionally, this can be implemented by having a read lock on the record or by validating if there is no newly committed version of the record before $T$ commits. (2)~\textit{Phantom avoidance}: the read set of $T$ should keep unchanged during the processing of $T$. Usually, this can be implemented by having an index/row-level lock on the entire read set or by re-scanning the read set in a validation phase.

However, to ensure SR by locks or re-scanning has a high cost, especially in a distributed setting. To avoid locks and re-scanning, we propose our own MV-OCC design with specific optimizations to maintain isolation. Our MV-OCC follows a general procedure of OCC but combines with our multi-version-based commit/abort rules.  Before we discuss the details, we first define the basic components and some necessary concepts.


\vspace{1mm}
\noindent{\bf Transaction Status Table (TST)} is a distributed table maintained in the RDMA-allocated memory in each server node. TST records the real-time status (i.e., \textit{execution}, \textit{validation}, \textit{commit}, \textit{abort}) of each active transaction $T$. Once $T$ starts the processing of a new phase, its worker thread will update $T$'s status in TST through RDMA atomic write. When another worker thread wants to check $T$'s status for its own transaction processing, it can apply one-sided RDMA read to fetch the value with microseconds latency.

\vspace{1mm}
\noindent{\bf Recent Committed Table (RCT)} is another distributed table but stored in the local memory of server nodes. RCT records the metadata of all active read-write transactions such as their IDs, commit time and write sets. RCT is indexed by a B-Tree using the commit time as the key. RCT will be used in the validation phase of each read-write transaction to avoid the high cost of read set re-scanning.

\vspace{1mm}
\noindent{\bf Multi-Versioning Read/Write Rules.} Figure~\ref{mvocc} shows the workflow of a transaction in \SysName{}. When a transaction $T$ reads a property object, it finds the visible version of the property value, i.e., the MV-tuple whose \textit{begin time} (\textit{BT}) and \textit{end time} (\textit{ET}) overlap with the BT of $T$. If $T$ wants to update the property object, it creates a new version, i.e., a new MV-tuple, and inserts its transaction ID (i.e., $TxID$) in the ET field of the current version and the BT field of the new version, as shown by the two MV-tuples in the Execution Phase in Figure~\ref{mvocc}. This indicates that the current property object is in the process of being updated, where the current version has been ``locked'' and the new version has not been committed yet. Then, if $T$ commits successfully later on, these two fields will be replaced by the \textit{commit time} (\textit{CT}) of $T$, i.e., $ct$($T$) in the Commit/Abort Phase in Figure~\ref{mvocc}. This indicates that the current version has ended at $ct$($T$) and a new version beginning at $ct$($T$) is created. But if $T$ aborts, then the BT field of the new version will be set as \textit{-Inf} to indicate that this version has become invisible forever. All the old versions whose ET is before the earliest BT of all active transactions, as well as the versions with \textit{-Inf} BT, will be cleaned and recycled back to the memory pool by garbage collection (\S\ref{impl}).

We now describe the details of the specific three phases of the transaction workflow in Figure~\ref{mvocc}.

\vspace{1mm}
\noindent{\bf In the execution phase}, a transaction $T$ first obtains its BT, i.e., the timestamp when its processing starts, for version visibility checking. Then, $T$ is executed and its read/write set is constructed by accessing the Vertex/Edge Tables and the MV-KVS, following the multi-versioning read/write rules. We propose an \textbf{optimistic pre-read} mechanism in the execution phase to reduce the abort rate. We illustrate the idea by the example given in the Execution Phase in Figure~\ref{mvocc}, where a transaction $T$ is doing the read-scanning on MV-tuples. Assume that $T$'s BT $> 10$. Then, $T$ will read the version (let it be V1) whose BT is 10 and ET is $TxID$ (meaning that another transaction $TxID$ is in the process of updating the corresponding property object). In this case, instead of aborting directly, optimistic pre-read assumes that $TxID$ will commit successfully later and executes $T$ according to the status of $TxID$ as follows. There are four possible cases:

\vspace{-1mm}
\begin{itemize}
	\setlength{\parskip}{0pt}
	\setlength{\itemsep}{0pt plus 1pt}
	\item $TxID$'s status is \textit{execution}: the version (let it be V2) next to V1 is a new version created by $TxID$ and V2 has not been committed yet. In this case, $T$ reads V1 and validates the read-set consistency (i.e., a new version is indeed not committed) in its validation phase.

	\item $TxID$'s status is \textit{validation}: we optimistically assume that $TxID$ will commit, and thus $T$ directly pre-reads V2 now but validates the commit dependency (i.e., if $TxID$ is indeed committed) in $T$'s validation phase. Note that reading V1 causes read instability and leads to abort.
	
	\item $TxID$'s status is \textit{commit}: $T$ directly reads V2 as $TxID$ has committed and the CT of $TxID$ is definitely earlier than the CT of $T$. 
	
	\item $TxID$'s status is \textit{abort}: $T$ ignores the new version V2, and reads the current version V1.

\end{itemize}	
\vspace{-1mm}

In the case of $T$ is a read-write transaction, if $T$ sees $TxID$ in the visible MV-tuples when $T$ is inserting a new version, it should abort itself immediately (except $T=TxID$), since this case belongs to a write-write conflict.

\vspace{1mm}
\noindent{\bf In the validation phase}, a read-write transaction $T$ first obtains its CT, which is the timestamp when the validation begins (i.e., when the transaction logically commits). Then, $T$ checks read stability and phantom avoidance through conflict checking. Specifically, based on the BT and CT of $T$, i.e., $bt$($T$) and $ct$($T$), we obtain from RCT all the read-write transactions, \textit{W-Trxs}, whose commit time falls in the period [$bt$($T$), $ct$($T$)] and their write sets may change the read set of $T$. If there is no overlapping element between the read set of $T$ and the write sets of \textit{W-Trxs}, then $T$ can commit successfully. Otherwise, $T$ should abort. The above conflict checking is executed as \textit{two set comparison} on all server nodes simultaneously in a MPP manner (\S\ref{exec}). We also propose an \textbf{optimistic validation} strategy to improve the success rate of commit. During the validation of $T$, if we find that $T$ is in conflict with another transaction $TxID$, where $TxID$ is in the validation phase too, we do not abort $TxID$ immediately. Instead, we optimistically assume that $TxID$ will abort later and continue the validation process of $T$ after recording such a dependency between $T$ and $TxID$ (i.e., $T$ should commit only if $TxID$ does abort).

At the end of the validation phase of $T$, we perform status checking for all the dependent transactions (if any). $T$ can commit itself only if all its dependent transactions due to optimistic pre-read have committed and all its dependent transactions due to optimistic validation have aborted. Note that if $T$ is a read-only transaction, its validation phase only needs to check those dependent transactions due to optimistic pre-read because $T$ has no write set.

\vspace{1mm}
\noindent{\bf In the commit/abort phase}, a read/write transaction $T$ physically effects its write set if $T$ commits, or discards if $T$ aborts, where the corresponding MV-tuples are updated according to the multi-versioning read/write rules.

\vspace{1mm}
\noindent  \textbf{Design Principle.} Following MV-OCC, there is no coarse-grained read/write locks or re-scanning during the entire transaction processing, which improves both the throughput and latency of transaction processing. The rationality of adopting the two optimistic optimizations in the execution and validation phases is that graph transaction workloads are mostly read-heavy as mentioned in~\S\ref{moti}. Moreover, to achieve snapshot isolation, MV-OCC can skip optimistic pre-read in the execution phase and the entire validation phase, and performs the regular commit (or abort when write-write conflict happens).

\subsection{Distributed Transaction Processing with MPP}\label{exec}
We now describe how \SysName{} leverages RDMA to process distributed transactions with the MV-OCC protocol.

With the use of RDMA, the design goal of \SysName's execution model is to effectively parallelize the transaction processing across servers while removing the computation bottlenecks (e.g., centralized coordinator, stragglers, locks). Existing RDMA-enabled databases~\cite{ChenWSCC16, WeiSCCC15, Myth} usually apply RDMA atomic verbs, compare-and-swap (CAS) and fetch-and-add (FAA), to lock and fetch records from remote machines to a local machine, and then perform the local transactional updates before writing them back to remote machines. However, such an approach is not efficient for graph transactions due to their large read/write sets. It is expensive to apply CAS \& FAA and locks on large amounts of data at a time, as this incurs high CPU overheads and impairs transaction throughput. Instead of fetching data remotely as~\cite{ChenWSCC16, WeiSCCC15} did, \SysName{} processes transactions on the servers where the data locates at and scales out the transaction processing across the servers through message passing via one-sided RDMA write. This approach avoids extra CPU overhead incurred by remote locking and RDMA-based CAS \& FAA by shifting the overhead to network, which is small by using Infiniband.

\begin{figure}[!t]
	\centering
	\includegraphics[width=0.99\columnwidth]{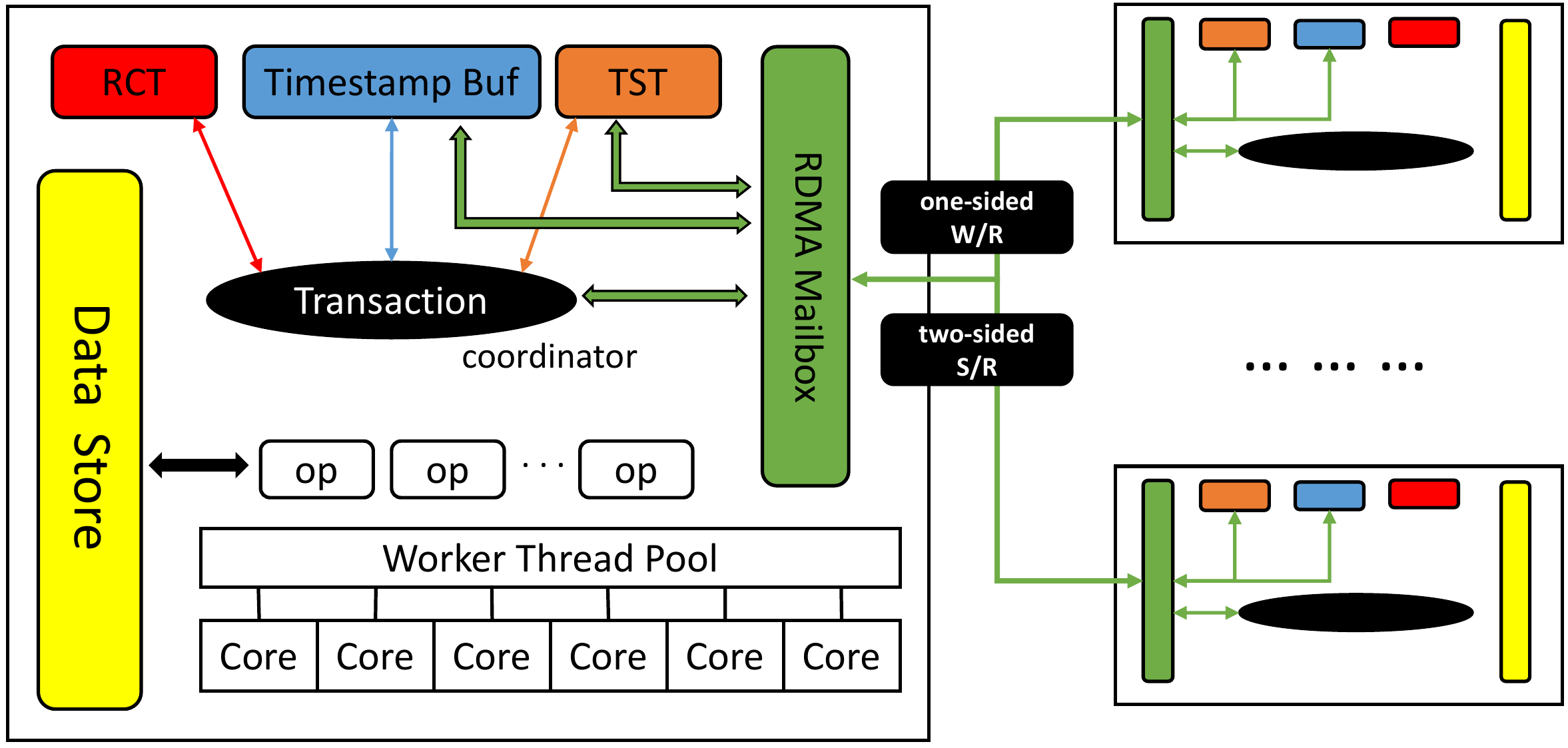}
	\vspace{-3mm}
	\caption{Distributed transaction processing in \SysName}\label{dtp}
	\vspace{-3mm}
\end{figure}

\SysName's decentralized architecture is shown as Figure~\ref{dtp}, where all servers have the same layout and play the same roles in the cluster. Without loss of generality, we only discuss the processing of one single transaction $T$ in one server. When a server receives $T$ (assigned by the cluster monitor), this server becomes the unique \textit{coordinator} of $T$ and handles the transactional metadata/result management of $T$, i.e., to aggregate intermediate/final results of $T$ if needed (e.g., for operators COUNT, MAX, etc.) and to update $T$'s status in the local shard of \textit{TST}. Both TST and RCT are read/write-enabled, shared data structures via RDMA. In order to maintain the atomicity and consistency of TST/RCT read/write, we adopt two different strategies according to their specific workloads.

Each update in TST is executed by a local atomic write on the coordinator server in order to avoid unsafe one-sided RDMA read from other server nodes. Applying atomic operation here is acceptable as a transaction status occupies only 2 bits. However, inserting an entire write set into RCT is a much bigger operation and implementing it on the RDMA allocated memory via RDMA-based CAS \& FAA for read/write consistency incurs high CPU overhead and thread contention. Thus, we maintain RCT in the local memory of the servers, while using two-sided RDMA send/recv to query and update the RCT entries in order to ensure its read/write consistency. In addition, we maintain a \textit{timestamp buf} in the RDMA allocated memory to synchronize the earliest BT among the active transactions on all servers through one-sided RDMA read, so as to obtain the global earliest BT, which is used by RCT to garbage-collect the expired entries (i.e., transactions and their write sets).

The entire processing of a transaction $T$ is split into shards cross all servers and executed in parallel. First, the coordinator of $T$ sends initial commands to all servers via one-sided RDMA write to trigger the start of $T$'s processing. But not every server can be activated for $T$, as it depends on the actual data locality of the read set of $T$. If a server indeed has one shard of $T$, then it follows the MV-OCC protocol to process $T$. According to the protocol, when $T$ needs to update its status, for example, a server has to abort $T$ due to a local conflict, it will synchronize $T$'s updated status to every shard of TST on the servers via one-sided RDMA write. Thus, at the beginning of each operation in $T$'s processing, the first step \SysName{} has to do is status checking. If $T$'s status has been changed, then we need to switch the processing of $T$ to the corresponding phase. In above example, $T$ will abort and terminate on all servers once they observe $T$'s new status. Thus, TST works as a global flag for each active transaction, and all servers can fast access/update it through RDMA. In contrast, the read/write sets of $T$ constructed in the execution phase of MV-OCC are recorded in each server locally. And in the validation phase of $T$, it first fetches all the potential conflicting transactions through RCT, and then does conflict checking on their read/write sets as described in~\S\ref{proto}.  Finally, if $T$ can commit successfully, besides following the procedures of MV-OCC in~\S\ref{proto}, the final result will be aggregated from each server to $T$'s coordinator and then sent to the client.

The MPP execution of each operator can speed up the query latency and improve the utilization of CPU resource. If one operator has a relatively higher load than the others, \SysName{} will use multiple threads to process this operation in parallel in order to achieve a shorter completion time. We follow the state-of-the-art MPP implementation from Grasper~\cite{ChenLFHCZHY19} while also take into consideration the side-effects of NUMA architecture~\cite{DrebesPHD016, NovakovicDBFG14}. As shown in Figure~\ref{dtp}, each thread in \SysName{}'s \textit{worker thread pool} is bound with one CPU core and further divided into several logical thread regions. This design can achieve better cache locality and memory locality on CPUs in cross-NUMA nodes. We omit the details here as this optimization is not the focus of our paper.
\section{System Implementation}\label{impl}
\SysName{} was implemented in C++, currently with 46K+ lines of code. To support RDMA communication, we developed a mailbox to achieve efficient RDMA one-sided read/write and two-sided send/recv based on RDMA \textit{ibverbs} library. We also provide a general TCP-based version of the system, i.e., \SysName{} without RDMA, which uses ZeroMQ TCP sockets to achieve point-to-point communication and MPI to coordinate the inter-process communication over Ethernet. Here we briefly discuss some of the system implementation details, while all code and details will be released on GitHub later.

\vspace{1mm}
\noindent \textbf{Timestamp Ordering.} Unified timestamp (together with MV-OCC) allows concurrent transactions to read a consistent snapshot of the database, while the timestamp order should be matched with the real time order. We follow the solution proposed by FaRMv2~\cite{RNCDNC19} for global time synchronization using Marzullo’s algorithm~\cite{MarzulloO85}, where any server in the cluster can play the role of clock master and other servers periodically synchronize their clocks via RDMA writes.

\vspace{1mm}
\noindent \textbf{Garbage Collection (GC).} MVCC-based protocol usually has a high overhead on the storage. Thus, as an in-memory graph database, GC is critical for \SysName{} to avoid the growth of memory consumption when servers run continuously. \SysName{}'s GC was implemented through one scanning thread and two GC threads as default, but users may configure the exact number of GC threads based on their workloads. However, recycling the memory slots occupied by obsolete objects requires write locks to guarantee data consistency for active read/write transactions, which leads to a degradation in transaction throughput and latency. In order to reduce the impact of GC, we separate the scanning process and execute GC jobs batch-by-batch. The scanning process collects the \textit{obsolete objects}, which include all old versions whose visible time periods have expired, all invalid versions that are generated due to the aborted transactions, and all empty Edge/VP/EP rows that have been deleted. We pack different types of obsolete objects into different types of GC tasks based on their costs, where each type of tasks handles only one type of obsolete objects and a specific threshold is set to control the batch size. The scanning thread periodically scans and collects obsolete objects. Once a batch has been collected, the GC threads will be activated to garbage-collect these objects. We show the importance of GC by experiments in~\S\ref{exp:gc}.

\vspace{1mm}
\noindent \textbf{Indexes.} Indexes are critical to achieve efficient query operations such as HAS and WHERE. \SysName{} supports standard indexes (e.g., hashtables, B+ trees) on text and numerical values for fast look-up or range search on vertex/edge labels and property values. Users can specify the type of indexes to be constructed and the specific target keys (e.g., a certain property) for indexing via a client console. Then, \SysName{} servers coordinate with each other to build a distributed index map in memory. When read-write transactions commit, the indexes will also be updated in real-time to maintain the consistency.
\section{Experimental Evaluation}\label{exp}
We compared \SysName{} with four popular graph databases, JanusGraph v.0.3.0~\cite{janusgraph}, ArangoDB v.3.6.2~\cite{arangodb}, Neo4J v.3.5.1~\cite{neo4j} and TigerGraph Developer Edition~\cite{TigerGraph}. The experiments were run on a cluster of 10 machines, each equipped with two 8-core Intel Xeon E5-2620v4 2.1GHz processors, 128GB memory and Mellanox ConnectX-3 40Gbps Infiniband HCA, running on CentOS 6.9 with OFED 3.2 Infiniband driver. For fair comparison, we used the same number of computing threads in each machine for all the systems, and tuned the configurations of each system to give its best performance as we could. All query latency reported are the average of five runs and all throughput values are averaged over 300 seconds.

\noindent{\bf Datasets.} We used four datasets: one small and one large synthetic property graphs created by LDBC-SNB\footnote{http://ldbcouncil.org/developer/snb} data generator, and two real-world property graphs crawled from DBpedia\footnote{https://wiki.dbpedia.org/downloads-2016-10} (including two parts, citation data and citation links) and Amazon Product\footnote{http://jmcauley.ucsd.edu/data/amazon/} (including product reviews and metadata). The small/large datasets were used for the system evaluation on the single-machine/distributed environment respectively. Table~\ref{datasets} lists the statistics of each dataset.

\begin{table}[!t]
	\centering\footnotesize
	\caption{Property graph datasets}\label{datasets}
	\vspace{-3mm}
	\begin{tabular}{|p{0.20\columnwidth-2\tabcolsep}
			p{0.20\columnwidth-2\tabcolsep}
			p{0.20\columnwidth-2\tabcolsep}
			p{0.20\columnwidth-2\tabcolsep}
			p{0.20\columnwidth-2\tabcolsep}|}
		\hline
		\textbf{Dataset} & $|V|$ & $|E|$ &  $|VP|$  & $|EP|$ \\  \hline
		LDBC-S & 23,850,377 & 139,854,135 & 153,761,078 & 37,769,010 \\ 
		DBPedia & 29,130,775 &	22,623,812 & 79,600,170	& 22,623,763 \\
		LDBC-L & 81,585,767 & 495,119,129 & 441,220,072 & 142,182,014 \\
		Amazon & 37,671,279 & 338,255,928 & 127,123,473 & 493,345,892 \\ \hline
	\end{tabular}
	\vspace{-3mm}
\end{table}

\noindent{\bf Query Benchmarks.} Lissandrini et al.~\cite{LissandriniBV18} proposed a benchmark (denoted as \textbf{LBV}) for graph database evaluation, which includes five categories of queries: \textit{Creation(C)}, \textit{Read(R)}, \textit{Update(U)}, \textit{Deletion(D)} and \textit{Traversal(T)}. \textit{R} and \textit{T} belong to \textit{READ} transactions, while \textit{C}, \textit{U} and \textit{D} belong to \textit{WRITE} transactions. We selected, with equal probability, most of the queries (i.e., [C:Q2-Q7], [R:Q11-Q15], [U:Q16-Q17], [D:Q18-Q21], [T:Q22-Q27]) in the benchmark (except few that are not for transactions, e.g., Q7-Q8 for counting all V/E) for throughput evaluation. In addition, we also selected 8 queries with heavier workloads, i.e., IC1-IC4 and IS1-IS4, in the \textbf{LDBC SNB} benchmark~\cite{ErlingALCGPPB15} for single-query latency evaluation. Note that we did not use the LDBC benchmark for the comprehensive evaluation as it does not support UPDATE/DELETE workloads and the queries are only applicable on its own synthetic datasets. We list the templates of all queries in both LBV and LDBC benchmarks in Appendix~\ref{appx:queries}.

\subsection{Evaluation of System Designs}\label{exp:opts}
We first evaluate the effectiveness of the various system designs, including the data store, the decentralized architecture, the optimizations in MV-OCC (i.e., optimistic pre-read and optimistic validation), and the speed-up due to RDMA-related designs.

\subsubsection{Evaluation of Individual System Designs}\label{exp:opts:individual}
To examine the effect of each individual design on the system performance, we created three variants of \SysName{}: \SysName{} without RDMA but using IPoIB (denoted as \textbf{IPoIB}), \SysName{} without the two optimizations in MV-OCC  (denoted as \textbf{No-Opt}), and a centralized version (denoted as \textbf{Cent}) with a global master (i.e., transaction coordinator).  We tested them on two large graphs, LDBC-L and Amazon, using 8 machines with 20 computing threads per machine. We used a mixed workload formed by READ and WRITE queries in the LBV benchmark. Half of the queries are WRITE, which create a relatively high-contention scenario.

\begin{figure}[!t]
	\centering
	\begin{subfigure}[h]{\columnwidth}
		\includegraphics[width=\linewidth]{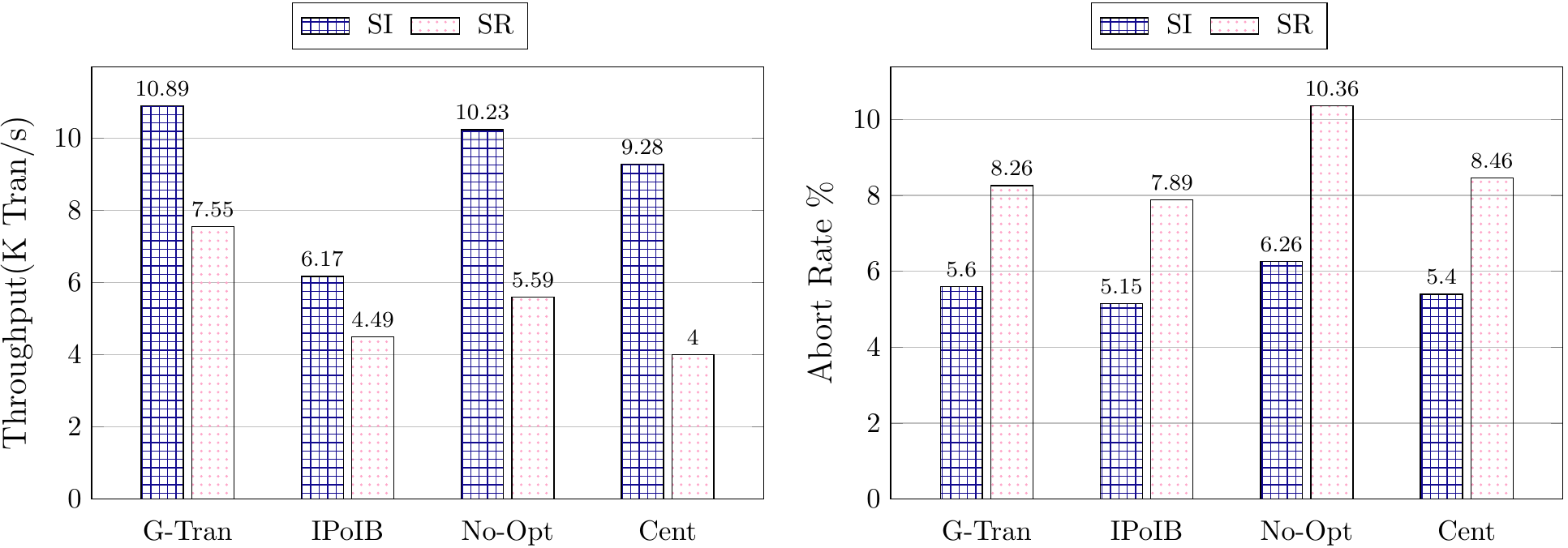}
		\vspace{-6mm}
		\caption{LDBC-L}
	\end{subfigure}
	\begin{subfigure}[h]{\columnwidth}
		\includegraphics[width=\linewidth]{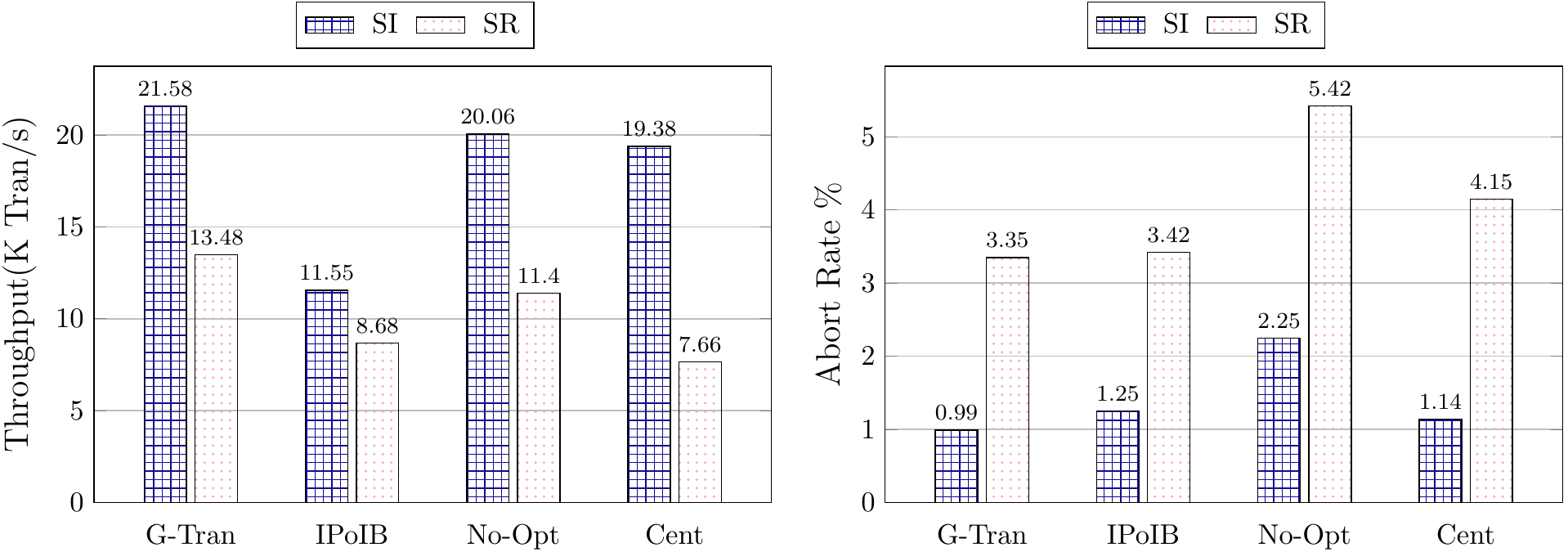}
		\vspace{-6mm}
		\caption{Amazon}
	\end{subfigure}
	\vspace{-4mm}
	\caption{The effects of individual system design}\label{effects}
	\vspace{-5mm}
\end{figure}

Figure~\ref{effects} reports the transaction throughput and the abort rate of \SysName{} and its three variants, at both SI and SR isolation levels. Compared with \SysName{}, the throughput of its IPoIB variant is reduced around 30\% - 50\%, which is due to the higher latency of normal network connection and the extra CPU overhead between NIC and the OS kernel.  However, the IPoIB variant still significantly outperforms existing systems as reported in~\S\ref{exp:comp}, which shows that other system designs also play important roles in \SysName{}'s high performance.

\begin{figure}[!t]
	\centering
	\begin{subfigure}[h]{0.49\columnwidth}
		\includegraphics[width=\linewidth]{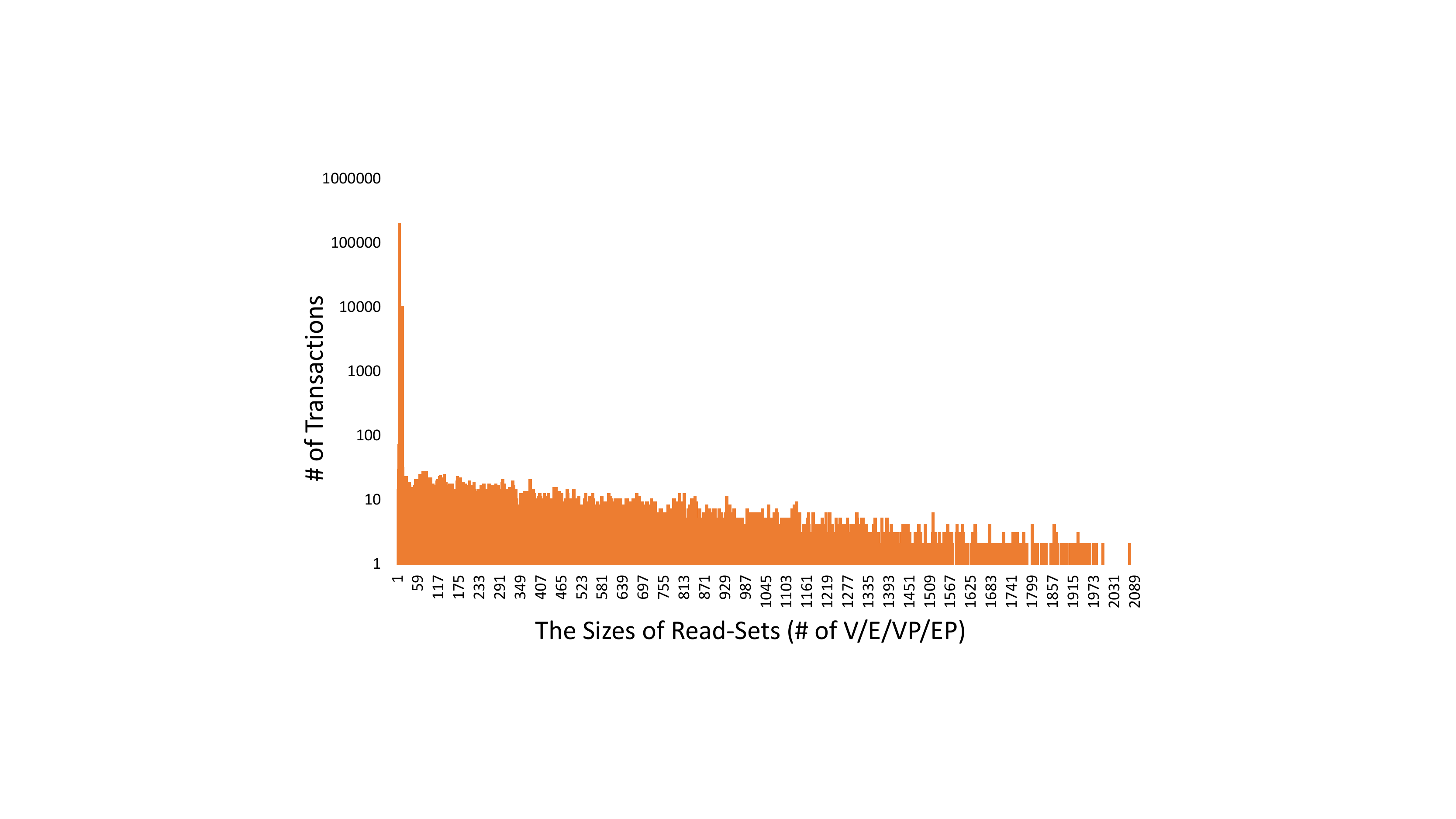}
		\vspace{-3mm}
	\end{subfigure}
	\hfill
	\begin{subfigure}[h]{0.49\columnwidth}
		\includegraphics[width=\linewidth]{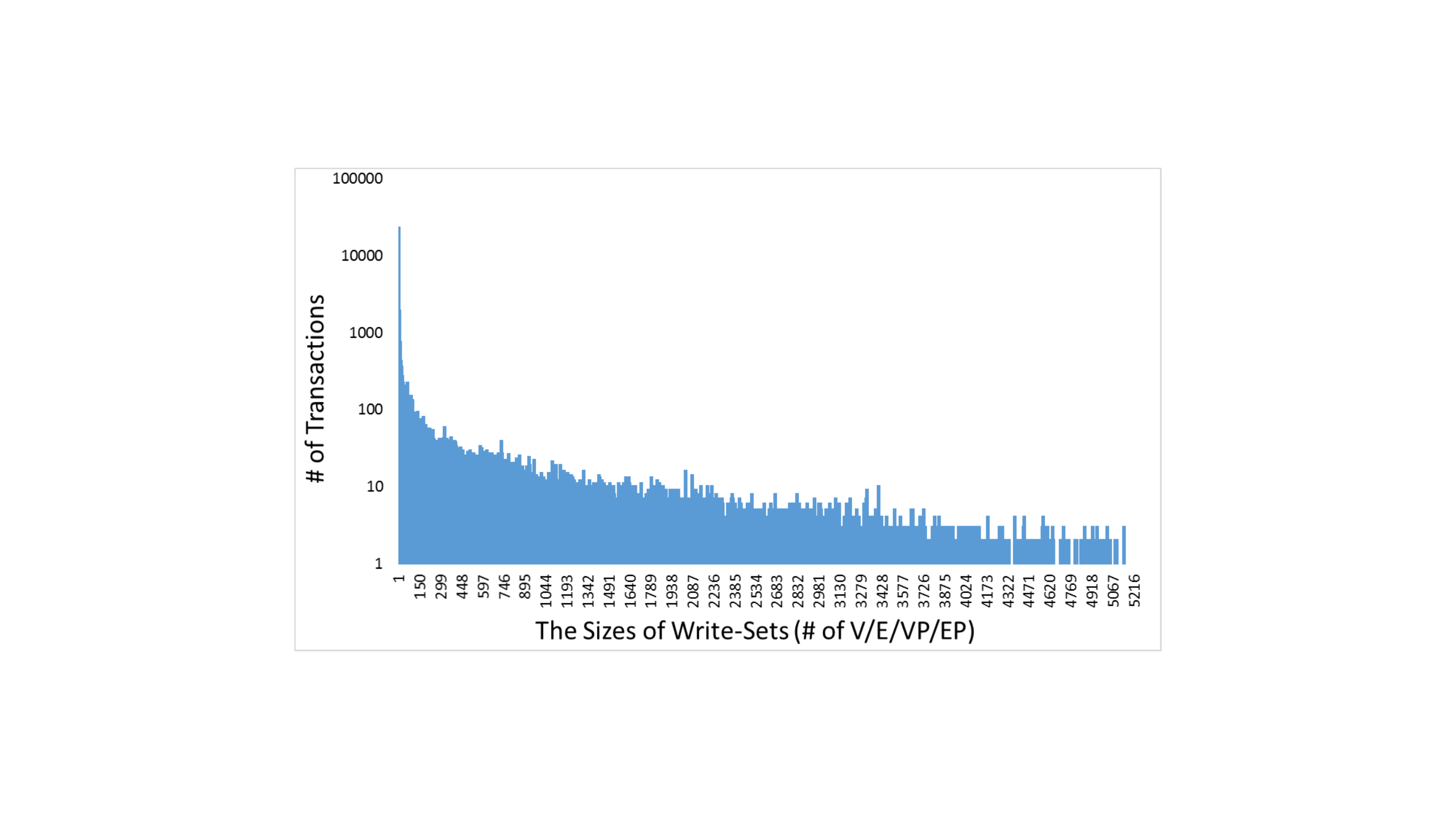}
		\vspace{-3mm}
	\end{subfigure}
	\vspace{-2mm}
	\caption{The distribution of the sizes of the read/write-sets of the LBV graph transactions on Amazon}\label{mixed-size}
	\vspace{-5mm}
\end{figure}

Disabling optimistic pre-read and optimistic validation in the MV-OCC protocol also leads to a degradation of the throughput, especially at SR isolation. As shown in Figure~\ref{effects}(a), on the LDBC-L dataset, No-Opt's throughput (4.49K tps) at SR drops 25.9\% compared with \SysName{}'s throughput (7.55K tps). No-Opt's throughput at SI is only slightly dropped as our multi-version-based solution can already eliminate much of the contention at SI isolation. However, the abort rate of the No-Opt variant is significantly higher. The increase in the abort rate is more obvious on the Amazon dataset as it is a real-world graph with a power-law distribution on vertex degree, leading to higher contention and abort rate. The result shows that the optimizations in the MV-OCC protocol can effectively improve the processing of concurrent graph transactions which have large read/write-sets. We report the distribution of the sizes of the read/write-sets of the transactions in Figure~\ref{mixed-size}, showing that although more transactions have relatively smaller read/write-sets, there are also a large number of transactions having large read/write-sets. This also explains the relatively high abort rate of \SysName{} at SR compared with that at SI.

To demonstrate the advantage of the decentralized architecture over RDMA for distributed transaction processing, we further compare \SysName{} with its Cent version. At SR isolation, Cent has the lowest throughput for both datasets (i.e, 4K tps and 7.66K tps). Because in Cent's setting, the master plays the role of a global coordinator which handles the coordinating tasks of all concurrent transactions. These tasks create significant CPU and network overheads on a single node, which becomes the bottleneck and limits the general processing power of the entire distributed system. We also observe an increase in the abort rate of the Cent version, because the averagely higher latency per transaction leads to higher contention and in turn increases the overall abort rate.

\subsubsection{Evaluation of Data Store}\label{exp:opts:store}
Next, we evaluate how the design of the data store effects \SysName{}'s performance.

Since we cannot disable the data store in \SysName{} individually as we did for the other features, we conducted the experiments on a single-machine setting to exclude the influence of \SysName{}'s RDMA-aware components and decentralized architecture. In addition, before we started to run the workloads, we first warmed up each system by running the mixed LBV workload in~\S\ref{exp:opts:individual} for an extended period of time, to simulate the real scenario that graph data locality has been broken after continuous updates. It leads to more random access on the entities of the graph, which can thus be used to indicate the effectiveness of the data layout.

We used 8 complex queries (i.e., IC1-IC4 and IS1-IS4) in the LDBC benchmark~\cite{ErlingALCGPPB15} to evaluate the performance. Consider these benchmark queries can only function on the LDBC synthetic dataset, we also involved a typical multi-hop query template into evaluation with the format:

\vspace{-3mm}
\begin{equation}
\textit{g.V().has([\textit{primary\_key}]).(both())$^k$}
\end{equation}

The \textit{both()} operator returns both the in-neighbors and out-neighbors of a source vertex. Here, \textit{both()} repeats $k=1,2,3,4$ times to represent a $k$-hop traversal from a starting vertex located by \textit{has()} operator with a given primary key (e.g., name).

We compared \SysName{} with JanusGraph, ArangoDB, Neo4J and TigerGraph, where all the systems ran on a single machine using 20 threads. Table~\ref{single1} and Table~\ref{single2} report the query latency of the LDBC queries and the $k$-hop traversal queries, respectively~\footnote{Note that JanusGraph failed to load DBPedia in 24 hours, and TigerGraph could not load DBPedia because it requires fixed-schema input while DBPedia has no schema as a real-world graph.}. \SysName{} achieves the shortest latency on all LDBC queries. In particular, for the complex queries, i.e., IC1-IC4, \SysName{}'s latency is two orders of magnitude smaller than that of JanusGraph and ArangoDB. The gap between \SysName{} and Neo4J is smaller but also 2-3 times in most cases. TigerGraph achieves competitive performance in its ``run'' stage, but every query needs an ``install'' stage before running and this ``install'' stage is costly. For the $k$-hop traversal queries, Table~\ref{single2} shows that latency on all systems increases exponentially as $k$ increases, because the size of the read-set grows exponentially for each hop of traversal on the graph. But after traversing more than 2 hops of neighbors, \SysName{} starts to show orders of magnitude advantage over the other systems.

\begin{table}[!t]
	\centering\footnotesize
	\caption{The latency (in \textit{msec}) of the LDBC queries}\label{single1}
	\vspace{-3mm}
	\begin{tabular}
		{|P{0.18\columnwidth-2\tabcolsep}
			P{0.11\columnwidth-2\tabcolsep}
			P{0.11\columnwidth-2\tabcolsep}
			P{0.1\columnwidth-2\tabcolsep}
			P{0.1\columnwidth-2\tabcolsep}
			P{0.1\columnwidth-2\tabcolsep}
			P{0.1\columnwidth-2\tabcolsep}
			P{0.1\columnwidth-2\tabcolsep}
			P{0.1\columnwidth-2\tabcolsep}|}
		\hlinew{1.2pt}
		\textbf{\scriptsize{LDBC-S}} & \textbf{IC1} & \textbf{IC2} & \textbf{IC3} & \textbf{IC4} & \textbf{IS1} & \textbf{IS2} & \textbf{IS3} & \textbf{IS4} \\ 
		\hlinew{1.2pt}
		\SysName{} 		& 7,085 	& 189 	& 4,986 	& 347 	& 0.5 & 13.1 & 4.6 & 0.4 \\
		Neo4J 			& 8,962 	& 824 	& 9.6E4 & 1,249 	& 1.1 & 25.4 & 6.8 & 1.6 \\	
		J.G. 		& 1.9E5 & 1.4E4 	& 1.3E6 & 1.3E5	& 1.2 & 20.6 & 2.7 & 0.9 \\	
		ArangoDB 		& 1.4E5 & 1,420	& 9.1E4 & 3,149 	& 1.1 & 58.9 & 33.6 & 0.8 \\	
		T.G.(install  	& 4.5E4 & 4.1E4 & 4.4E4 & 4.5E4 & 3.8E4 & 3.9E4 & 3.6E4 & 3.5E4 \\
		+ run) 			&  +63.5 &  +19.1 &  +370 &  +21.3 &  +8.2 &  +15.2 &  +8.9 &  +6.7 \\
		\hlinew{1.2pt}
	\end{tabular}
	\vspace{-1mm}
\end{table}

\begin{table}[!t]
	\centering\footnotesize
	\caption{The latency (in \textit{msec}) of $k$-hop traversal queries}\label{single2}
	\vspace{-3mm}
	\begin{tabular}{|p{0.20\columnwidth-2\tabcolsep}
			p{0.20\columnwidth-2\tabcolsep}
			p{0.20\columnwidth-2\tabcolsep}
			p{0.20\columnwidth-2\tabcolsep}
			p{0.20\columnwidth-2\tabcolsep}|}
		\hlinew{1.2pt}
		\textbf{DBpedia} & \textbf{Q1} & \textbf{Q2} &  \textbf{Q3} & \textbf{Q4} \\
		\hlinew{1.2pt}
		\SysName{} 		& 0.9 	& 9.7 		& 966 		& 8,084 \\ 
		Neo4J 			& 1.8 	& 20.5 		& 1,128 	& 19,217  \\
		ArangoDB 		& 24.4 	& 93.5	& 17,659 	& 287,012 \\
		\hlinew{1.2pt}
		\textbf{LDBC-S} & \textbf{Q1} & \textbf{Q2} &  \textbf{Q3} & \textbf{Q4} \\
		\hlinew{1.2pt}
		\SysName{} 		& 0.3 	& 349 		& 5,338 		& 42,452 \\ 
		Neo4J 			& 1.4 	& 758 		& 9,911 	& 128,762  \\
		J.G.  			& 1.3 	& 661	& 42,916 	& 1,211,714 \\
		ArangoDB 		& 0.6 	& 2,006	& 36,715 	& >4h \\
		T.G.(install  	& 40,518 & 59,968 	& 92,770 	& 132,766 \\
		+ run) 			&  +8.82 &  +389 	&  +3,788 	&  +62,053 \\
		\hlinew{1.2pt}
	\end{tabular}
	\vspace{-2mm}
\end{table}

We explain the results by analyzing the storage design of each system as follows. Neo4J stores all edges of the entire graph into a global sequential table. Consequently, a graph traversal will suffer from jump addressing in physical storage space on Neo4J, because each newly inserted edge is appended directly to the tail of the sequential table without locality guarantee. As for TigerGraph, although it is not open source, considering that it requires users to indicate the edge type (e.g., friendship) during a traversal, we conjecture TigerGraph stores edges separately based on their types. Then, this layout will require an extra aggregation to merge those separate edge sets when a traversal involves multi-type edges, which explains why TigerGraph has poor performance on the $k$-hop traversal queries. Both JanusGraph and ArangoDB store edges as individual cells or collections. Specifically, JanusGraph stores edge cells of one vertex together with its property cells in one row, which incurs extra overhead for locating edges from properties. ArangoDB stores edges as collections of documents. Thus, the execution of multi-hop traversal needs to locate the connected edges of each vertex by searching on edge collections. In comparison, \SysName{} divides arbitrary-length adjacency-lists into fixed-size rows, which provides more efficient edge insertion and deletion through allocating new rows or compacting under filled rows. Note that we allocate a new row only after filling up all blanks in the existing row and the sparse row compaction is processed in periodical. As a result, traversals in \SysName{} can access the edges of a vertex sequentially and the data locality will not be broken even after frequent updates.

\subsection{Throughput Analysis}\label{exp:comp}
Now, we compared \SysName{} with existing systems in terms of throughput in both distributed and single-machine settings. We applied three workloads for evaluation: (1)~\textit{Read-Only} (\textit{RO}), formed by READ queries in the LBV benchmark, (2)~\textit{Read-Intensive} (\textit{RI}), which consists of 80\% READ and 20\% WRITE queries, (3)~\textit{Write-Intensive} (\textit{WI}), which consists of 20\% READ and 80\% WRITE queries. The specific query values (e.g., VPIDs, EPIDs, etc.) in the query templates were randomly generated to form the query-sets for throughput evaluation.

\subsubsection{Distributed Processing}\label{exp:distributed}
We first compared \SysName{} and its IPoIB variant (\SysName{}-IPoIB) with JanusGraph (using HBase as the storage backend) and ArangoDB. For Neo4J and TigerGraph, we only can download their Developer Edition which does not support distributed processing. We set isolation at SI because both ArangoDB and JanusGraph do not support SR.

\begin{figure}[!t]
	\centering
	\begin{subfigure}[h]{0.49\columnwidth}
		\includegraphics[width=\linewidth]{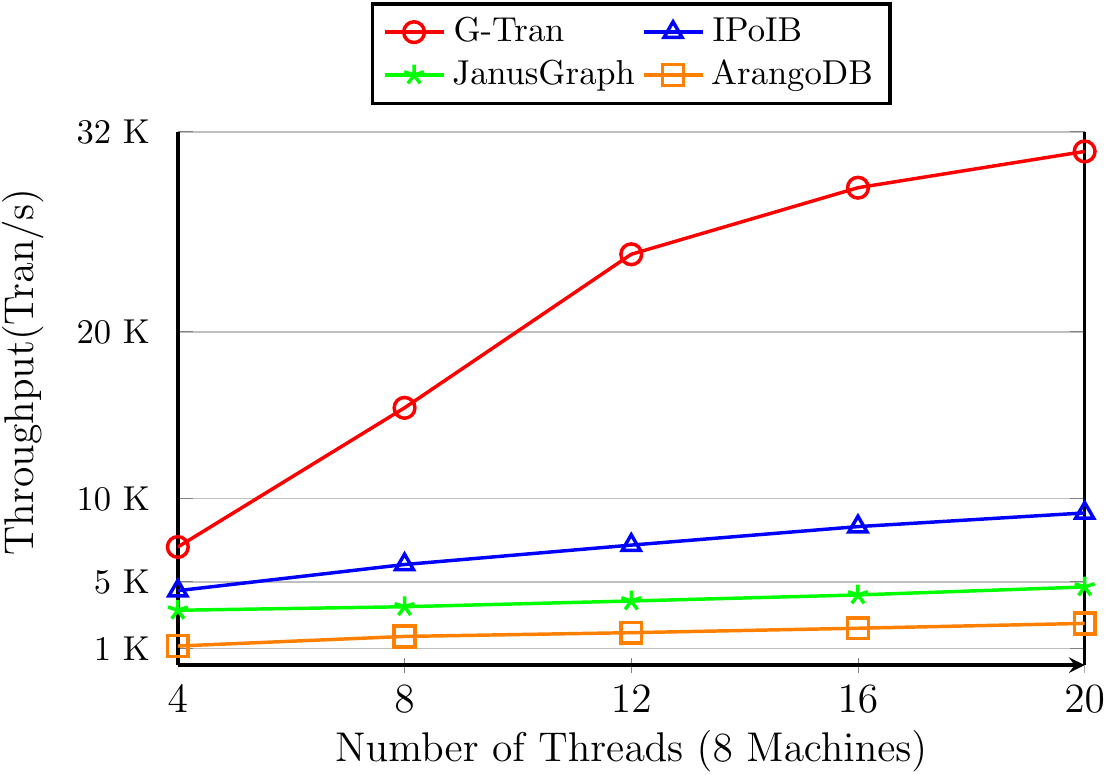}
	\end{subfigure}
	\hfill
	\begin{subfigure}[h]{0.49\columnwidth}
		\includegraphics[width=\linewidth]{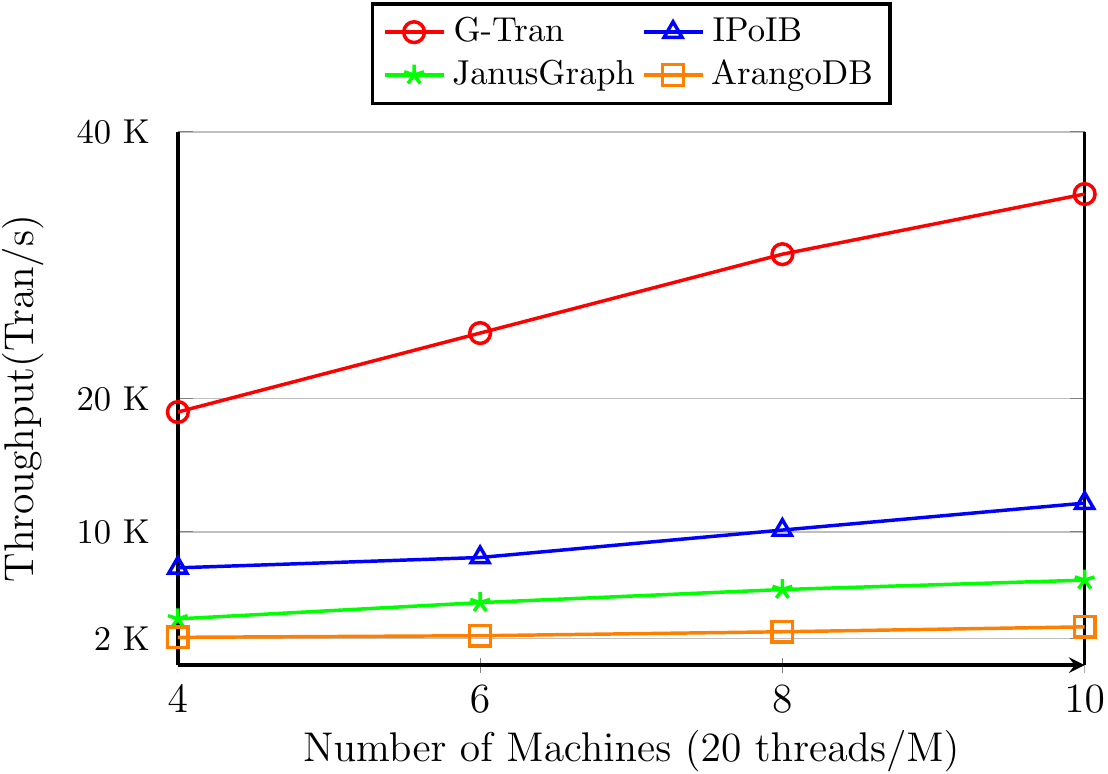}
	\end{subfigure}
	\vspace{-3mm}
	\caption{Scale-up and scale-out throughput on Read-Only workload over the LDBC-L dataset}\label{scale-ro}
	\vspace{-1mm}
\end{figure}

\begin{figure}[!t]
	\centering
	\begin{subfigure}[h]{0.49\columnwidth}
		\includegraphics[width=\linewidth]{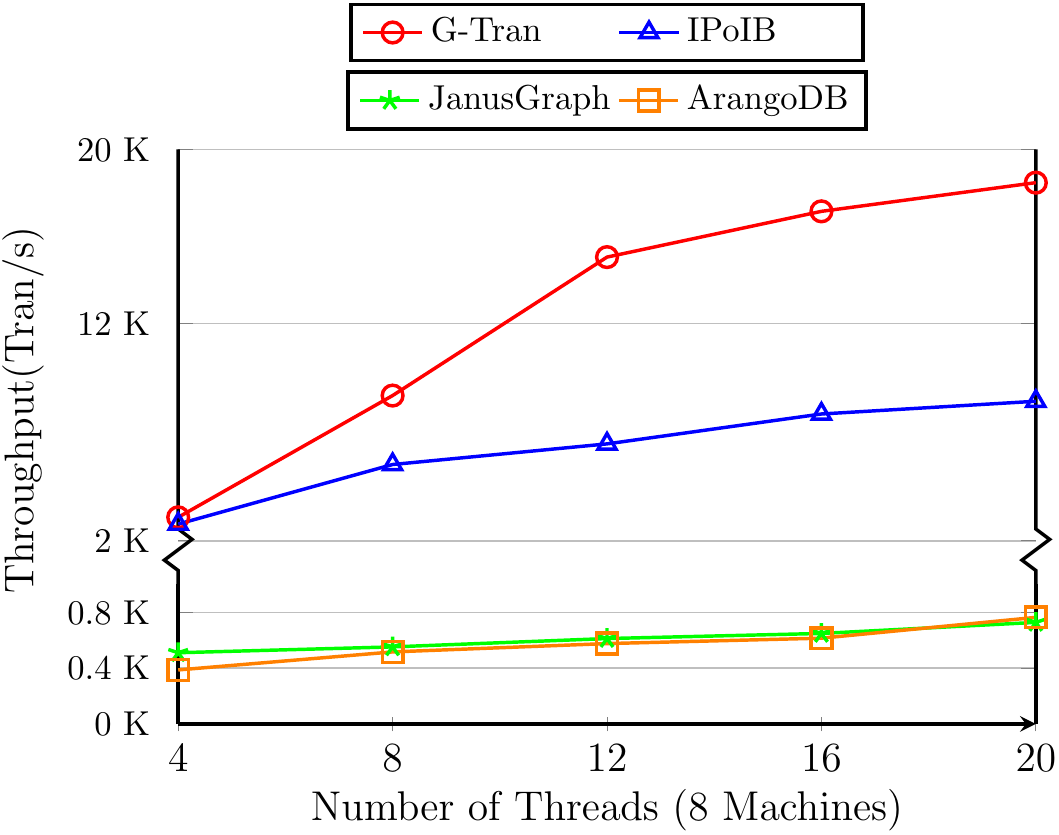}
	\end{subfigure}
	\hfill
	\begin{subfigure}[h]{0.49\columnwidth}
		\includegraphics[width=\linewidth]{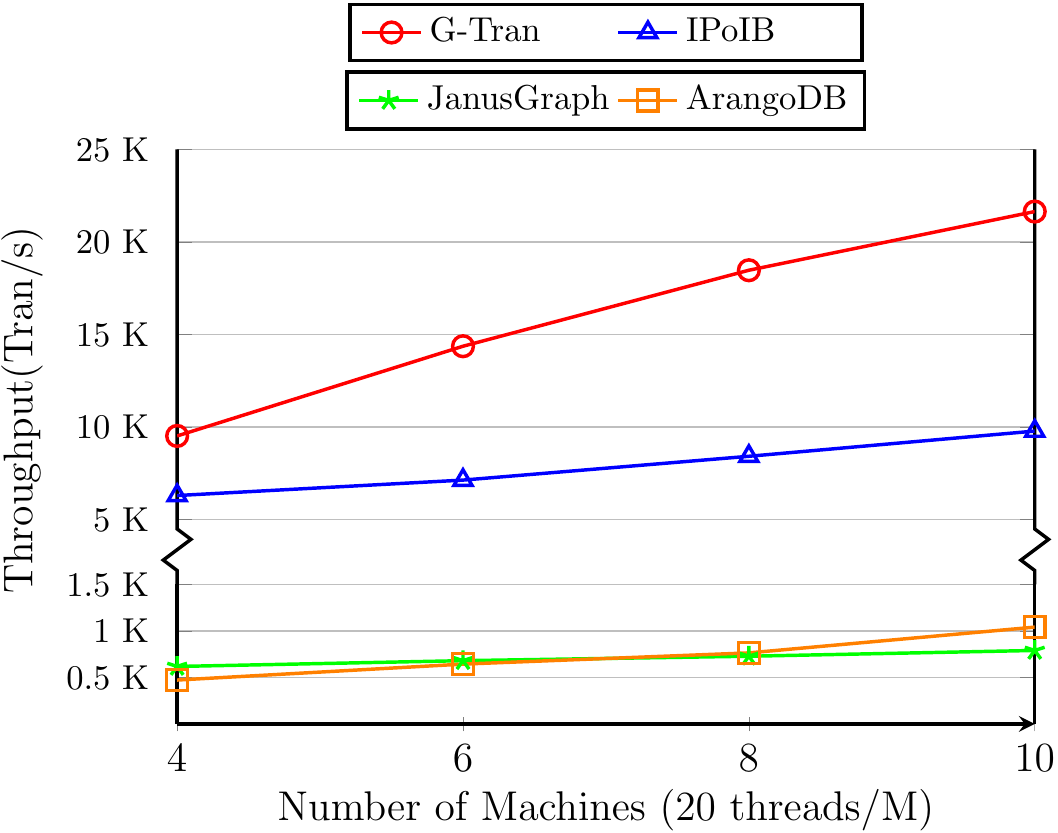}
	\end{subfigure}
	\vspace{-3mm}
	\caption{Scale-up and scale-out throughput on Read-Intensive workload over the LDBC-L dataset}\label{scale-ri}
	\vspace{-1mm}
\end{figure}

\begin{figure}[!t]
	\centering
	\begin{subfigure}[h]{0.49\columnwidth}
		\includegraphics[width=\linewidth]{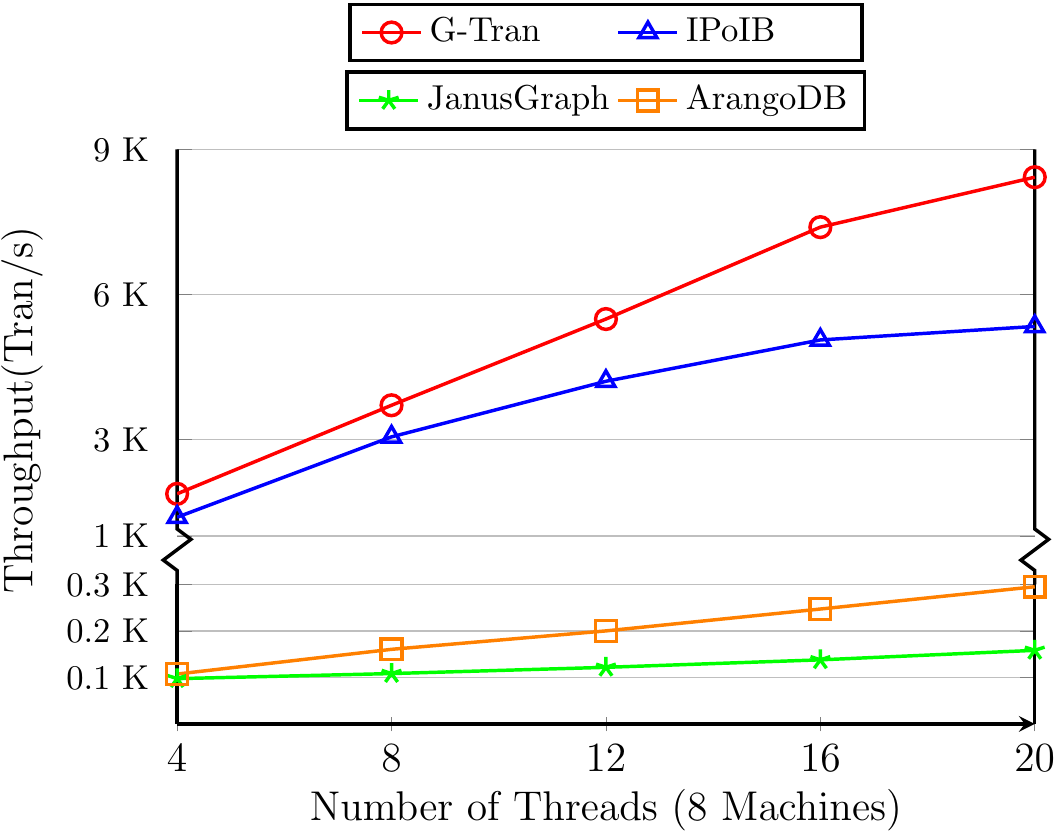}
	\end{subfigure}
	\hfill
	\begin{subfigure}[h]{0.49\columnwidth}
		\includegraphics[width=\linewidth]{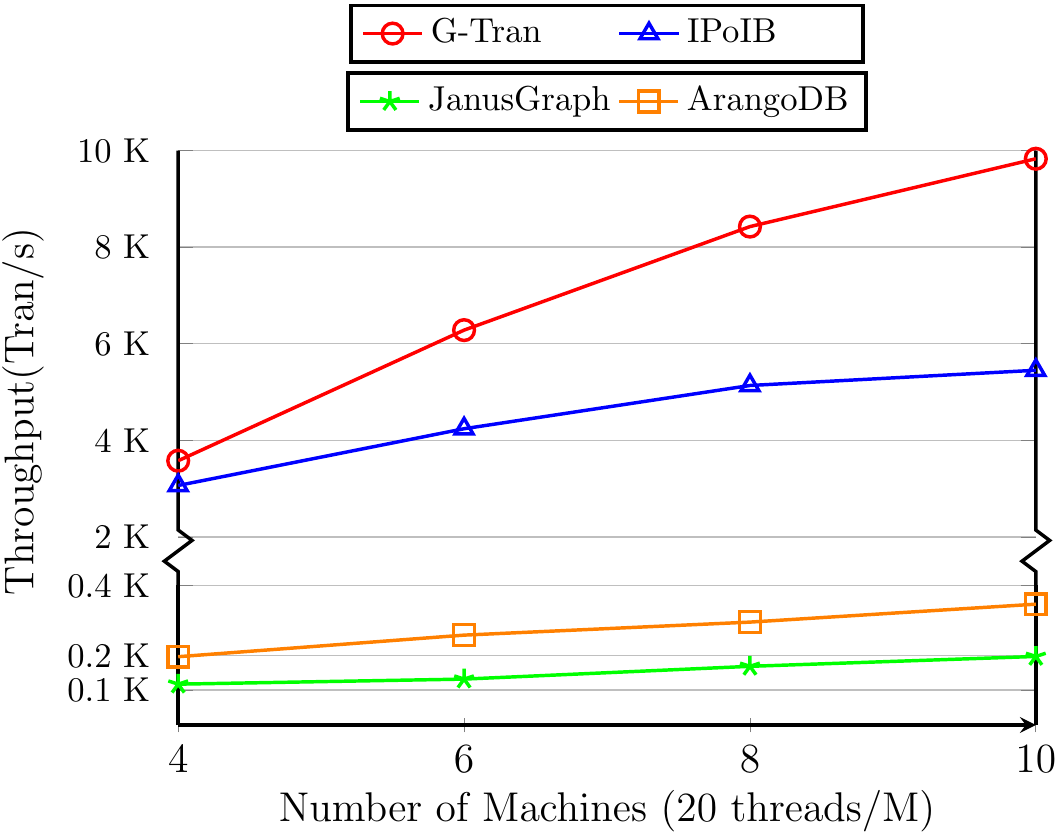}
	\end{subfigure}
	\vspace{-3mm}
	\caption{Scale-up and scale-out throughput on Write-Intensive workload over the LDBC-L dataset}\label{scale-wi}
	\vspace{-1mm}
\end{figure}

Figures~\ref{scale-ro}-\ref{scale-wi} report the throughput scalability of each system on the three workloads respectively over the LDBC-L dataset. Specifically, for scale-up evaluation, we depolyed all systems on 8 machines but varying the number of threads per machine from 4 to 20. For scale-out evaluation, we ran them on 4 to 10 machines with 20 threads/machine. Similar performance pattern is also observed on Amazon dataset, we skip the detailed reports here due to the limited space.

\SysName{} achieves significantly higher throughput than other systems over all three workloads. Although the performance of \SysName{}-IPoIB is degraded when RDMA features are disabled, it still outperforms both JanusGraph and ArangoDB in all cases. This indicates that the use of RDMA in \SysName{} is not the only reason for its high performance, but other system components are also important as discussed in~\S\ref{exp:opts}. As we increase the number of machines from 4 to 10, the results show that using RDMA indeed brings an advantage to distributed processing because \SysName{} has higher rate of increase in throughput than \SysName{}-IPoIB. Overall, when more machines are used, \SysName{}'s throughput increases more on RI and WI workloads. This phenomenon can also be observed in scale-up performance. In contrast, the throughput of both JanusGraph and ArangoDB are relatively low and scales poorly. Besides caused by their data storage as we discussed in~\S\ref{exp:opts:store}, this is also related to how the execution engine of each system processes the concurrent transactions. \SysName{}'s MPP model enables higher parallelism to process each transaction when more resources are available. Thus, these transactions can commit earlier and there is less contention when accessing data. However, both JanusGraph and ArangoDB follow the one-thread-one-transaction mechanism. When more threads are available, more transactions will join in the system and being processed simultaneously. This increases the contention among all concurrent transactions and incurs more overheads. Moreover, when more machines are involved, the graph will be partitioned into more shards, which breaks the locality of the graph and leads to extra overhead on communication.

\subsubsection{Single-Machine Processing}\label{exp:single}
We compared \SysName{} with Neo4J, TigerGraph, JanusGraph and ArangoDB for single-machine transaction throughput at both SI and SR. We used BerkeleyDB as the backend of JanusGraph for its evaluation at SR, as only BerkeleyDB supports SR as a single-machine system. Neo4J and TigerGraph do not support SI, while ArangoDB does not support SR. TigerGraph and JanusGraph failed to load DBPedia as explained in~\S\ref{exp:opts:store}. We used 20 threads for all the systems.

\begin{figure}[!t]
	\centering
	\begin{subfigure}[h]{0.49\columnwidth}
		\includegraphics[width=\linewidth]{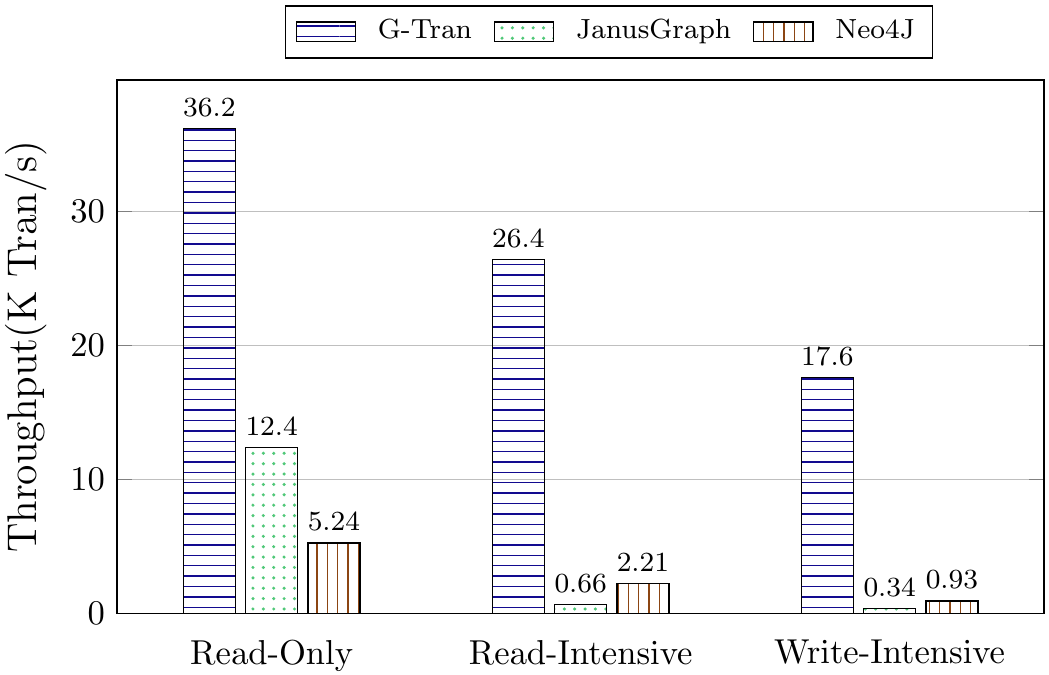}
		\vspace{-6mm}
		\caption{DBPedia}
	\end{subfigure}
	\hfill
	\begin{subfigure}[h]{0.49\columnwidth}
		\includegraphics[width=\linewidth]{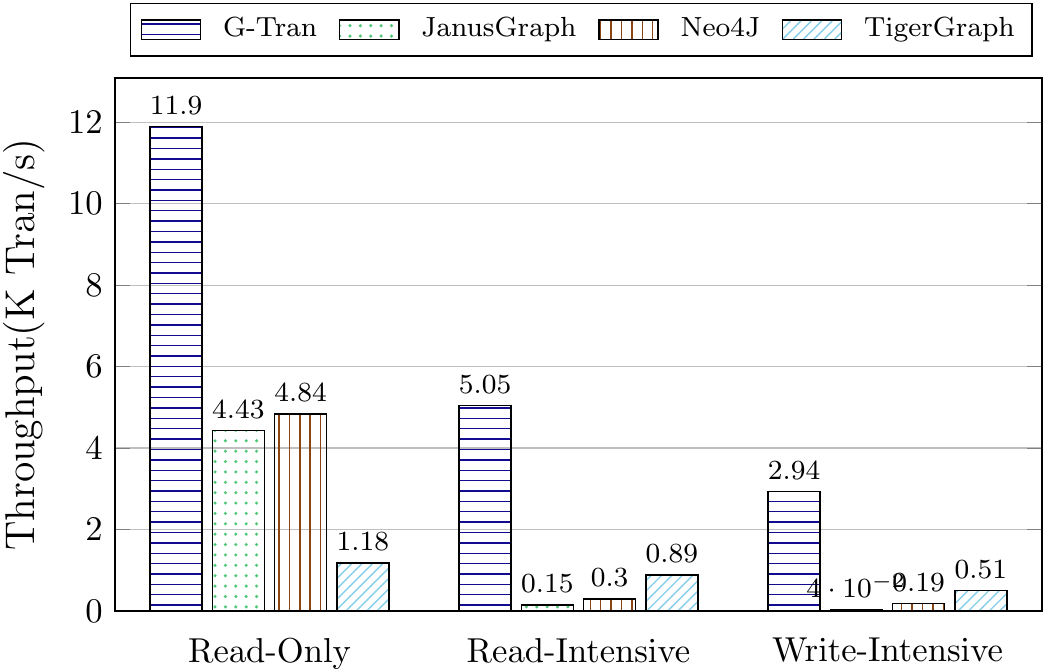}
		\vspace{-6mm}
		\caption{LDBC-S}
	\end{subfigure}
	\vspace{-3mm}
	\caption{Single-machine throughput at SR}\label{single-tp-SR}
	\vspace{-1mm}
\end{figure}

\begin{figure}[!t]
	\centering
	\begin{subfigure}[h]{0.49\columnwidth}
		\includegraphics[width=\linewidth]{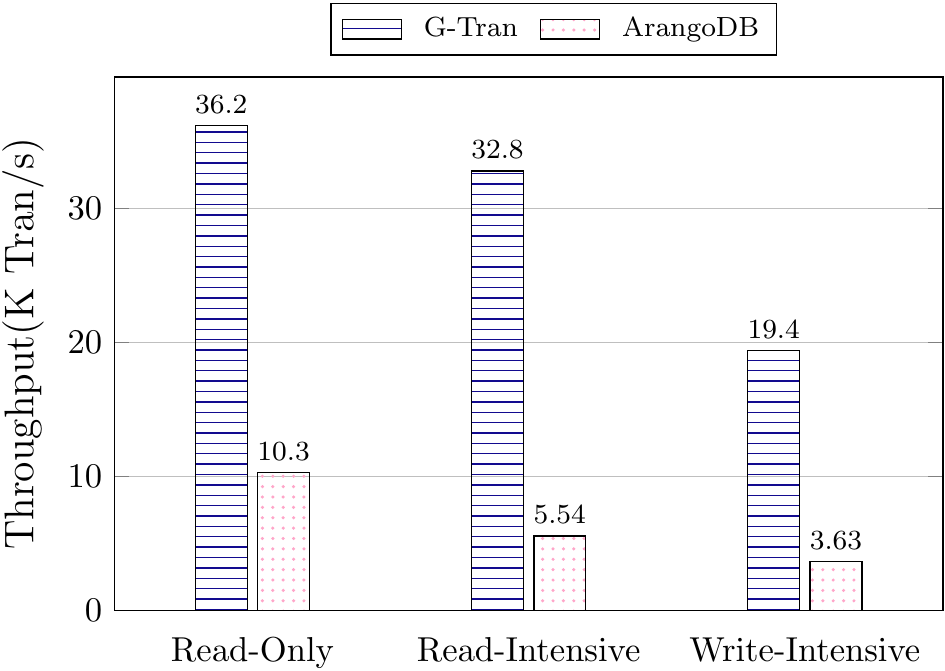}
		\vspace{-6mm}
		\caption{DBPedia}
	\end{subfigure}
	\hfill
	\begin{subfigure}[h]{0.49\columnwidth}
		\includegraphics[width=\linewidth]{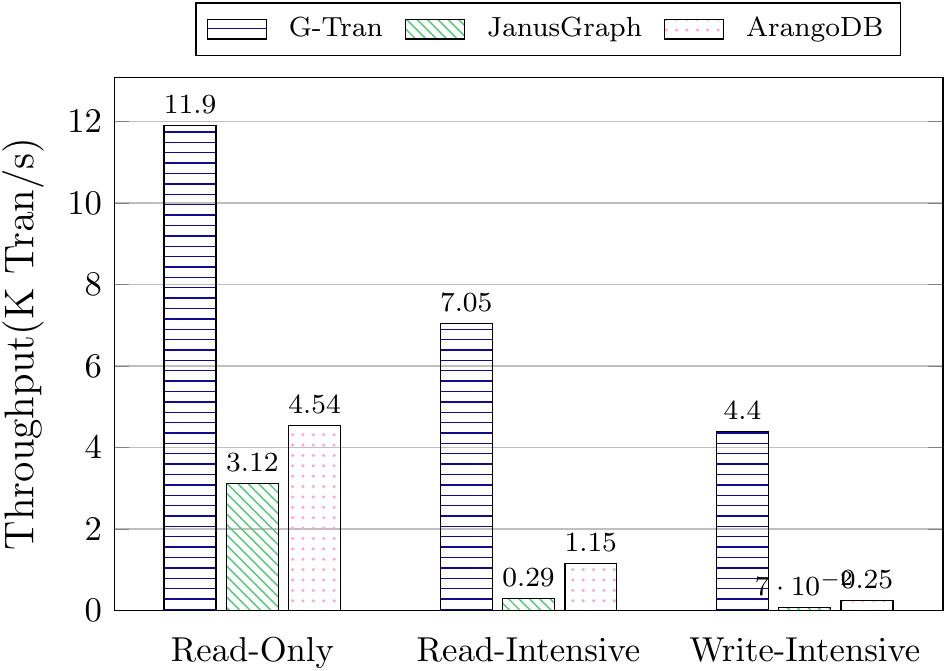}
		\vspace{-6mm}
		\caption{LDBC-S}
	\end{subfigure}%
	\vspace{-3mm}
	\caption{Single-machine throughput at SI}\label{single-tp-SI}
	\vspace{-1mm}
\end{figure}

As reported in Figure~\ref{single-tp-SR} and Figure~\ref{single-tp-SI}, \SysName{} achieves significantly higher throughput than other systems on all workloads at both SR and SI. Note that in single-machine setting, the performance advantages of \SysName{} do not come from RDMA and its decentralized architecture, but mainly from its data store design and MV-OCC protocol. JanusGraph has worse performance than others because \SysName{}, TigerGraph and Neo4J have native graph stores with tailored designs for transaction processing, while JanusGraph is built upon a general NoSQL-based store. \SysName{}'s good performance comes mainly from (1)~its multi-version based storage, which enables lock-free snapshot reads; (2)~the optimizations in MV-OCC protocol, which reduces contention among the concurrent transactions with less abort and retry; (3)~its MPP execution engine, which enables parallel processing inside each transaction. By contrast, Neo4J's transaction engine does not support SR natively but requires explicit locks on query language level to achieve SR. TigerGraph's low throughput is related to its costly ``install'' stage, which is used to pre-translate and optimize queries before their real execution. \SysName{} requires neither explicit lock nor ``install'' stage to process transactions. In addition, ArangoDB can achieve higher throughput than JanusGraph due to its MVCC-based storage. But it has inefficient data layout on graph store, thus ArangoDB's throughput is still lower than that of \SysName{}.

\subsection{Performance of Garbage Collection}\label{exp:gc}
We also evaluated the effectiveness of \SysName's Garbage Collection (GC) mechanism by comparing the performance difference between enabling and disabling GC during regular transaction processing. We conducted the experiment using 8 machines with 20 computing threads per machine on the LDBC-L dataset and executed the same workload at SR isolation as we did in~\S\ref{exp:opts:individual}.

\begin{figure}[!t]
	\centering
	\includegraphics[width=0.97\columnwidth]{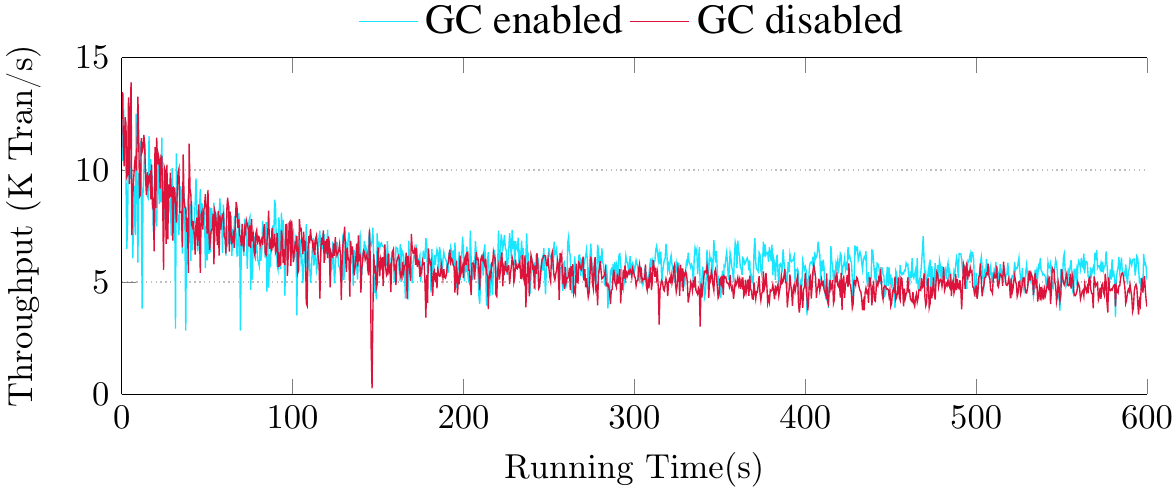}
	\vspace{-4mm}
	\caption{Throughput with GC enabled/disabled}\label{gc-throughput}
	\vspace{-1mm}
\end{figure}

\begin{figure}[!t]
	\centering
	\includegraphics[width=0.97\columnwidth]{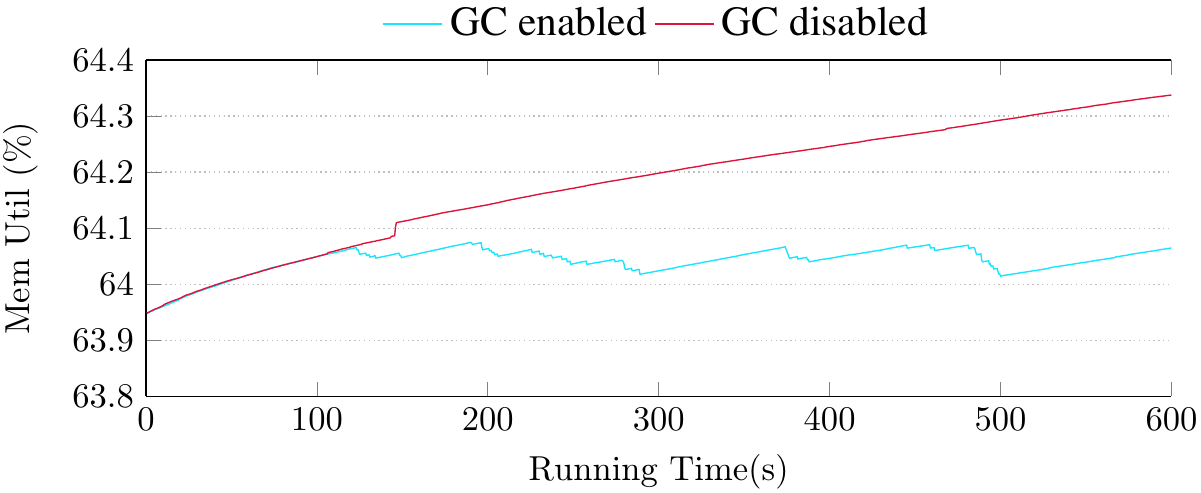}
	\vspace{-3mm}
	\caption{Memory consumption with GC enabled/disabled}\label{gc-memory}
	\vspace{-1mm}
\end{figure}

Figure~\ref{gc-throughput} and Figure~\ref{gc-memory} report the real-time throughput and memory utilization of \SysName{} for a period of 600 seconds from the beginning, i.e., as soon as \SysName{} finishes the data loading. The throughput is higher at the beginning as there is not many GC jobs to do, and the system becomes stabilized at around 200 seconds. The result shows that, when GC is disabled, the memory consumption of \SysName{} increases linearly with time. But when GC is enabled, the memory consumption remains relatively stable. We observe some obvious drops at 200s, 295s, 380s and 500s in Figure~\ref{gc-memory}, which are as a result of the periodical garbage collection that releases the occupied memory and returns it back to the memory pool. During the whole process, the negative side-effects of GC execution on transaction throughput is minor. We can compare the two curves in Figure~\ref{gc-throughput}, the throughput of \SysName{} in the GC-enabled case follows very closely to the case when GC is disabled, showing that our GC mechanism has low overhead. Actually, at the later period, in the GC-enabled case, \SysName{}'s throughput even has a tiny increase. This is because at this time, the system storage has been accumulated by many invalid/expired versions of various objects, to clean them up timely can de-fragment those sparse rows in the vertex/edge tables and accordingly improve the entire memory locality, which helps improve the efficiency of data scan and search.

As the GC mechanism categorizes and packages different GC objects into different types of tasks, the result demonstrates the effectiveness of \SysName{}'s GC mechanism for recycling various types of obsolete objects in a timely fashion (as otherwise the memory footprint would increase linearly over time). We can also observe a significant drop on the throughput in the GC-disabled curve at round 140s in Figure~\ref{gc-throughput}. While at the same time point, the GC-disabled curve for memory utilization in Figure~\ref{gc-memory} shows a non-continuous increase. We guess that at this moment, there were a large number of write transactions coming into the system, and system itself suffered from their processing and commits/aborts in short-term.
\section{Related Work}\label{otherwork}
\noindent{\bf Graph Databases.} There exist many graph databases like Neo4J~\cite{neo4j}, Titan~\cite{titan}, JanusGraph~\cite{janusgraph}, ArangoDB~\cite{arangodb} and OrientDB~\cite{orientdb}, etc. However, they have no MPP-based distributed transaction processing and most of them support only low isolation level (e.g., \textit{snapshot}, \textit{read committed}). Both Grasper~\cite{ChenLFHCZHY19, ChenWDHLLC20} and TigerGraph~\cite{TigerGraph} target at massive graph queries with native graph store, but Grasper focuses only on OLAP workload instead of OLTP and TigerGraph is not designed for high performance. A1~\cite{BuragohainRBCCC20} is an in-memory graph database built upon FaRM~\cite{DragojevicNCH14, NNRSB15}, while it has no graph native design on memory management and transactional optimizations. Compared with these existing systems, \SysName{} has distinguished designs in its graph-native data store, RDMA-featured decentralized architecture and MV-OCC, which contribute to its high performance as reported in~\S\ref{exp}.

\vspace{0.2mm}
\noindent{\bf Graph Processing Systems.} Many graph processing systems have been proposed~\cite{ZhangCYCLB17, avery2011giraph, chen2015powerlyra, gonzalez2012powergraph, gonzalez2014graphx, salihoglu2013gps, wu2015g, ZhuCZM16, YanHLCCWZ18, YanCCLB19} based on Pregel model~\cite{malewicz2010pregel} or other computation models~\cite{tian2013think, yan2014blogel, ZhangACACLL19}. But they focus on offline graph workloads such as PageRank, Connected Component and SSSP. There are also other systems that aim at complex graph analytics and mining~\cite{ChenLZYYC18, ChenWHFHLC19, abs-1709-03110, YanCCCS18}, or streaming graph processing~\cite{MariappanV19, DhulipalaBS19, ChenWZYC16}, which can not be used for graph OLTP.

\vspace{0.2mm}
\noindent{\bf Distributed Transaction Processing.} Many systems have been proposed in recent years for distributed transaction processing, including Google's Spanner~\cite{TWW13}, Granola~\cite{CowlingL12}, FaSST~\cite{KaliaKA16} and others~\cite{FILMSVZ13, PorobicLTA14, ThomsonDWRSA12, TuZKLM13}. Some of them leverage new hardwares such as RDMA, HTM and NVM to achieve high performance (e.g., FaRM~\cite{DragojevicNCH14, NNRSB15} and DrTM~\cite{ChenWSCC16, WeiSCCC15}). FaRM proposed an RDMA-friendly protocol to enable strict serializiable transactions with high throughput, low latency, and high availability. DrTM proposed an OCC protocol combining both HTM~\cite{BrownA16} and RDMA to ensure the strong consistency and atomicity. However, these systems are not specially designed for graph, which has its own unique challenges (\S\ref{intro} and~\S\ref{moti}).
\section{Conclusions}\label{con}

We presented \SysName{}, a high-performance RDMA-based graph database with both serializability and snapshot isolation support. To tackle the unique challenges of graph transaction processing, we used RDMA one-sided/two-sided verbs to reduce system overheads from network and CPUs. \SysName{} also proposed its own data layout, transaction protocol and decentralized architecture to achieve an overall good performance in terms of both latency and throughput. For future work, we plan to enhance \SysName{} on durability and availability by addressing fault tolerance and data replication.

\bibliographystyle{ACM-Reference-Format}
\bibliography{gtran}

\clearpage
\appendix
\section{Benchmark Query Templates}\label{appx:queries}

We list the benchmark queries used in our experimental evaluation in~\S\ref{exp}. Table~\ref{template1} lists the query templates for the \textbf{LBV} benchmark~\cite{LissandriniBV18} and Table~\ref{template2} lists the query templates for the \textbf{LDBC SNB} benchmark~\cite{ErlingALCGPPB15}. The values of the vertex/edge id and properties are randomly sampled from the respective datasets on which the queries are evaluated.

\begin{table}[!t]
	\centering\footnotesize
	\caption{Query templates in the \textbf{LBV} benchmark}\label{template1}
	\vspace{-3mm}
	\begin{tabular}{|p{0.1\columnwidth-2\tabcolsep}
			|p{0.9\columnwidth-2\tabcolsep}|}
		\hline
		Q2 & \tabincell{l}{
			g.addVertex(p[]) // create a new vertex with properties p
		} \\  \hline
		Q3 & \tabincell{l}{
			g.addEdge(v1, v2, l)  // add an edge with label $l$ from $v1$ to $v2$
		} \\  \hline
		Q4 & \tabincell{l}{
			g.addEdge(v1, v2, l, p[])
		} \\  \hline
		Q5 & \tabincell{l}{
			v.setProperty(Name, Value)
		} \\  \hline
		Q6 & \tabincell{l}{
			e.setProperty(Name, Value)
		} \\  \hline
		Q7 & \tabincell{l}{
			g.addVertex(...); g.addEdge(...) \\
			 // add a new vertex, and then edges to the vertex
		} \\  \hline
		Q11 & \tabincell{l}{
			g.V.has(Name, Value) // Vertices with property $Name=Value$
		} \\  \hline
		Q12 & \tabincell{l}{
			g.E.has(Name, Value)
		} \\  \hline
		Q13 & \tabincell{l}{
			g.E.has("label", l)
		} \\  \hline
		Q14 & \tabincell{l}{
			g.V(id) // The vertex with identifier $id$
		} \\  \hline
		Q15 & \tabincell{l}{
			g.E(id)
		} \\  \hline
		Q16 & \tabincell{l}{
			v.setProperty(Name, Value)
		} \\  \hline
		Q17 & \tabincell{l}{
			e.setProperty(Name, Value)
		} \\  \hline
		Q18 & \tabincell{l}{
			g.removeVertex(id)
		} \\  \hline
		Q19 & \tabincell{l}{
			g.removeEdge(id)
		} \\  \hline
		Q20 & \tabincell{l}{
			v.removeProperty(Name)
		} \\  \hline
		Q21 & \tabincell{l}{
			e.removeProperty(Name)
		} \\  \hline
		Q22 & \tabincell{l}{
			v.in() // Vertices adjacent to $v$ via incoming edges
		} \\  \hline
		Q23 & \tabincell{l}{
			v.out()
		} \\  \hline
		Q24 & \tabincell{l}{
			v.both("l") // Vertices adjacent to $v$ via edges labeled $l$
		} \\  \hline
		Q25 & \tabincell{l}{
			v.inE.label.dedup() // Labels of incoming edges of $v$ (no dupl.)
		} \\  \hline
		Q26 & \tabincell{l}{
			v.outE.label.dedup()
		} \\  \hline
		Q27 & \tabincell{l}{
			v.bothE.label.dedup()
		} \\  \hline
	\end{tabular}
\end{table}

\begin{table}[!h]
	\centering\footnotesize
	\caption{Query templates in the \textbf{LDBC SNB} benchmark}\label{template2}
	\vspace{-3mm}
	\begin{tabular}{|p{0.1\columnwidth-2\tabcolsep}
			|p{0.9\columnwidth-2\tabcolsep}|}
		\hline
		IS1 & \tabincell{l}{
			g.V().has("ori\_id", "ID").union( \\
			\qquad properties("firstName", "lastName", "birthday", \\ \qquad \qquad	"locationIP", "browserUsed"), \\
			\qquad out("isLocatedIn").properties("name") \\
			)
		}  \\  \hline
		IS2 & \tabincell{l}{
			g.V().has("ori\_id", "ID").in("hasCreator") \\
			.order("creationDate", decr).limit(10).as("msg").union( \\
				\qquad hasLabel("post").out("hasCreator").properties("firstName"), \\
				\qquad hasLabel("comment").out("replyOf").out("hasCreator")\\
				\qquad \qquad .properties("firstName")\\
			).as("author").select("msg", "author")
		}  \\  \hline
		IS3 & \tabincell{l}{
			g.V().has("ori\_id", "ID").union( \\
			\qquad outE("knows").as("knowEdge1").inV().as("friend1")\\
			\qquad \qquad .select("knowEdge1").properties("creationDate")\\
			\qquad \qquad .as("creationDate1").select("friend1", "creationDate1"), \\
			\qquad inE("knows").as("knowEdge2").outV().as("friend2")\\
			\qquad \qquad .select("knowEdge2").properties("creationDate").\\
			\qquad \qquad as("creationDate2").select("friend2", "creationDate2")\\
			)
		}  \\  \hline
		IS4 & \tabincell{l}{
			g.V().has("ori\_id", "ID").union(\\
			\qquad hasLabel("comment").properties("content", "creationDate"), \\
			\qquad hasLabel("post").properties("imageFile", "creationDate")\\
			)
		}  \\  \hline
		IC1 & \tabincell{l}{
			g.V().has("ori\_id", "ID").as(`a').both("knows").as(`b').\\
			both("knows").where(neq(`a')).as(`c').\\
			both("knows").where(neq(`a')).as(`d').\\
			union( select(`b'), union(select(`c'), select(`d')))\\
			.has("Key", "Value").as("person").union(\\
			\qquad union(\\
				\qquad \qquad out("isLocatedIn").properties("name"), \\
				\qquad \qquad out("studyAt").out("isLocatedIn").properties("name") \\
				\qquad ), \\
			\qquad  out("workAt").out("isLocatedIn").properties("name")\\
			).as("place").select("person", "place")
		}  \\  \hline
		IC2 & \tabincell{l}{
			g.V().has("ori\_id", "ID").both("knows").in("hasCreator").\\
			has("creationDate", lte("Date")).\\
			order("creationDate", decr).limit(1).properties()\\
		}  \\  \hline
		IC3 & \tabincell{l}{
			g.V().has("ori\_id", "ID").as(`a').union(\\
			\qquad both("knows"), \\
			\qquad both("knows").both("knows") \\
			).where(neq(`a')).dedup().not(\\
			\qquad out("isLocatedIn").out("isPartOf").or(\\ 
			\qquad \qquad has("CountryKey", "Country1"), \\
			\qquad  \qquad has("CountryKey", "Country2"))\\
			).as("person").in("hasCreator").\\
			has("creationDate", between("StartDate", "EndDate")).\\
			or(\\
			\qquad out("isLocatedIn").has("$CountryKey", "$Country1"),\\ 
			\qquad out("isLocatedIn").has("$CountryKey", "$Country2")\\
			).select("person").groupCount()
		}  \\  \hline
		IC4 & \tabincell{l}{
			g.V().has("ori\_id", "ID").both("knows").in("hasCreator").\\
			hasLabel("post").has("creationDate", \\
			\qquad between("StartDate", "EndDate")\\
			).out("hasTag").not(\\
			\qquad in("hasTag").hasLabel("post").\\
			\qquad has("creationDate", lte("Date"))\\
			).groupCount("name")
		}  \\  \hline
	\end{tabular}
\end{table}

\end{document}